\documentclass[onecolumn]{IEEEtran}
\usepackage{amsthm}
\usepackage{amsmath}
\usepackage{amssymb}
\usepackage{esint}
\usepackage{color}

\theoremstyle{plain}
\newtheorem{thm}{\protect\theoremname}
\theoremstyle{definition}
\newtheorem{example}[thm]{\protect\examplename}
\theoremstyle{plain}
\newtheorem{lem}[thm]{\protect\lemmaname}
\theoremstyle{definition}
\newtheorem{defn}[thm]{\protect\definitionname}

\providecommand{\definitionname}{Definition}
\providecommand{\examplename}{Example}
\providecommand{\lemmaname}{Lemma}
\providecommand{\theoremname}{Theorem}

\begin{document}

\title{Minimal Evacuation Times and Stability}

\maketitle

\begin{abstract}
We consider a system where packets (jobs) arrive for processing using one of the policies in a given class. We study the connection between the minimal evacuation times and the stability region of the system under the given class of policies. The result is used to establish the equality of information theoretic capacity region and system stability region for the multiuser broadcast erasure channel with feedback. 
\end{abstract}

\section{Introduction}

In this work we consider a time slotted system where packets 
arrive to one of $n$ different input queues - there may be other
system queues to which packets are placed during their processing.
The packets are processed by a policy from an admissible class. We study the connection between system stability
and minimal \emph{evacuation time}, i.e. the time it takes to complete
processing a number of packets placed at the input queues at time 0, provided that no further arrivals occur afterwards.
Under certain general assumptions on admissible policies and system
statistics, it is shown that the stability region of the system is
completely characterized by the asymptotic growth rate of minimal
evacuation time. We make very few assumptions on the system structure
and hence the result is applicable to a large number of applications
in communications as well as more general control
systems. However, we point out that the result, while intuitive, has
to be applied with caution since there are systems for which its application
leads to wrong conclusions. As an application to our methodology,
we consider the $N\mbox{-user}$ broadcast erasure channel with feedback.
In this setup, we compare the information theoretic capacity region
with the stability region and show that they are equal.

Concepts akin to evacuation time and their relation to stability
have been investigated in earlier works. Baccelli and Foss \cite{BacFoss}
consider a system fed by a marked point process and operating under
a given policy. The concept of \emph{dater} is used to describe the
time of last activity in the system, if the system is fed only by
the $m$th to $n$th , $m\leq n$ of the points of the marked process.
Assuming that the dater is a deterministic function of the arrival
times and the marks of the point process, and under additional assumption
on dater sample paths, they show that stability under the specified
policy is characterized by the asymptotic behavior of daters. These
results are extended to continuous time input processes by Altman
\cite{altman94}. In our setup, the system evolution may depend on
random factors as well as the characteristics of the arrival process.
Moreover, we do not make sample path assumptions on specific policies.
We rather specify features that admissible policies may have, and
based on these we characterize the stability region of the class of
admissible policies by the asymptotic growth rate of minimal (over
all admissible policies) evacuation times.

A different, yet related, methodology is developed by Meyn \cite{meyn};
the \emph{workload} $w(t)$ is defined as the time the
server must work to clear all of the inventory of the system at time
$t$ when operating in the fluid limit. This basic concept is elaborated
and used to derive significant results and obtain intuition for good
control policies in specific complex networks. The concept of workload
is closely related to the evacuation time, however we make minimal
assumptions on system structure and the derived results are applicable
to more general systems. 

Regarding the relation between the information theoretic capacity
and queueing theoretic stability regions, the equality of these has been shown recently in \cite{effros} for systems without
feedback. The system studied in this work uses feedback, and as will
be seen it can be derived in a simple manner based on stability characterization
through evacuation time.

\subsection{Preliminaries}
In the following, we use the vector notation $\boldsymbol{x}=\left[x_{1},x_{2},...,x_{n}\right]^{\top}.$
Also $\boldsymbol{x}\geq\boldsymbol{y}$ means $x_{i}\geq y_{i},\: i=1,2...,n$
and
\[
\left\lceil \boldsymbol{x}\right\rceil \doteq\left[\left\lceil x_{1}\right\rceil ,...,\left\lceil x_{n}\right\rceil \right],
\]
where $\left\lceil x\right\rceil $ is the least integer larger than
or equal to $x.$ With $\boldsymbol{m},\:\boldsymbol{k}$ we denote
vectors with nonnegative integer coordinates and with $\boldsymbol{r},\:\boldsymbol{s}$
vectors with nonnegative real number coordinates. 

\section{System Model and Admissible Policies\label{sec:System-Model}}

We consider a time-slotted system where slot $t=0,1,...$ corresponds
to the time interval $[t,\: t+1).$ The system has $n$ input queues
of infinite length where packets\footnote{In this work we use the term \emph{packet}, that describes an arriving unit in a communication network. However, our work applies to any general service system with arrival processes and queues, e.g. manufacturing systems, road networks, network switches, etc. Therefore, the subsequent discussion and results should be understood generically.} arrive. Packets arriving at each input may have certain properties,
e.g., service times, priorities, routing options, etc. There may be additional queues in the system, where
packets may be placed during its operation. At the beginning of time
slot $t$, i.e., at time $t$, $A_{i}\left(t\right)$ packets arrive
at input $i$. (In particular, we use $A_i(0)$ to denote the
number of packets in the queue of input $i$ when the system commences operation
at $t=0$.) We assume
that the arrival processes satisfy the ergodicity condition

\begin{equation}
\lim_{t\rightarrow\infty}\frac{\sum_{\tau=0}^{t}A_{i}\left(\tau\right)}{t}=\lambda_{i}>0,\: i=1,2,...,n\label{eq:1.0}
\end{equation}
as well as, 
\begin{equation}
\lim_{t\rightarrow\infty}\frac{\mathbb{E}\left[\sum_{\tau=1}^{t}A_{i}\left(t\right)\right]}{t}=\lambda_{i},\: i=1,2,...,n\label{eq:-1}
\end{equation}

The operation of the system is characterized by a finite set
of \emph{system states} $\mathcal{S}$, and control sets $\mathcal{G}_{s}$ for
each $s\in\mathcal{S}$: if at the beginning of a slot the system
state is $s\in\mathcal{S}$, one of the available controls $g\in\mathcal{G}_{s}$
is applied. There may be randomness in the behavior of the system, that
is, given $s$ and $g$ at the beginning of a slot, the system state
and the results at the end of a slot (e.g. packet erasures) may be random. 
For example, this makes the model particularly useful in wireless
networks, where outcomes of transmissions may depend on channel state and
ambient noise.

Arriving packets are processed by the system following a policy $\pi$,
belonging to a class of admissible policies $\Pi.$ At time $t,$
when the system state is $s,$ an admissible policy specifies:
\begin{enumerate}
\item The control $g\in\mathcal{G}_{s}$ to be chosen. 
\item An action $\alpha$ among a set of available actions $\mathcal{A}_{g}$
when control $g$ is chosen. An action specifies how packets are handled
within the system.
\end{enumerate}
The choice of controls and actions depends on the ``system history''
up to $t,$ denoted by $\mathcal{H}_{t}$. The history $\mathcal{H}_{t}$ includes
all information about packet arrival instants, packet departure instants,
system states, controls, actions taken and results, up to and including
time $t$.

Note that in the mathematical analysis
of systems, the ``state'' of the mathematical model may include
part of $\mathcal{H}_{t}$, and actions are usually not distinguished
from controls. For the purposes of this work, the terms
\emph{system states} and \emph{controls}
are explicitly used to refer to the operational characteristics of the system,
and are distinct from the history $\mathcal{H}_{t}$ and actions taken
once the system characteristics are set. For example, the
sizes of the queues at time $t$ should usually be considered as part of the
information captured by $\mathcal{H}_{t}$, rather than the system state, unless
the queue states directly impact the set of controls available to the system.
Also, we emphasize that
the choice of one action or another within a given control (for example, which
particular packet is transmitted from a given queue) 
does not affect the system state or slot outcome.
This distinction is needed
in order to define well the statistical assumptions needed for the
development that follows. We next present several examples to clarify these
notions. 

\medskip{}

\begin{example}
\label{Example-preemptive}Assume a wireless transmitter which can
transmit to a destination over one of two transmission channels, I or
II (e.g. over two different carriers). Data arriving at the transmitter
is classified in two types A, B. Packets from each of the classes
are placed in distinct infinite size queues.

The channels can be in one of four states, $\left(s_{1},s_{2}\right)\in\left\{ (l,h),(h,l),(l,l),(h,h)\right\} $.
The controls $\mathcal{G}_{s}$ available when in state $s=(s_{1},s_{2})$
determine a) the channel to be used for transmission, and
b) the transmission power $p$. This choice determines
the rate or transmission $r(p,s)$ in packets per second over the
chosen channel. Once a control $g$ is chosen, the action set $\mathcal{A}_{g}$
consists of two elements, $a_{A}$ and $a_{B}$ indicating the type
of data to be transmitted over the chosen channel. The choice of action does
not make a difference to the dynamics of the system state.
\end{example}
\medskip{}

\begin{example}
\label{Example-states}Consider a communication system consisting
of two nodes, $a$, $b$. Arriving packets are stored in an infinite
queue at node $a$ and must be delivered to node $b.$ The two nodes
are connected with two links, $\ell_{1},\:\ell_{2},$ \emph{at most
one} of which may be activated at a time. If link $\ell_{1}$ is activated,
a packet can be successfully transmitted in one slot, but both links
cannot be activated for the next $9$ slots. If link $\ell_{2}$ is
activated, a transmitted packet is erased with probability .5 (and
received successfully with probability .5) and both links can be activated
in the next slot. 

The states for this system can be described by the elements $\left\{ 0,1,2,...,9\right\}$, where
state $0$ means that both links can be activated and state $i\geq1$
means that no link can be activated for the next $i$ slots.

The control set for state $0$ is $\mathcal{G}_{0}=\left\{ g_{0},\: g_{1},\: g_{2}\right\} $
where $g_{0}$ means no link activation, $g_{1}$ means activation
of link $\ell_{1}$ and $g_{2}$ means activation of link $\ell_{2}$$.$
The control set for the rest of the states consist only of $g_{0}$.
From state $0,$ if control $g_{0}$ or $g_{2}$ is taken, the state
returns to $0$ in the next slot, while if $g_{1}$ is taken the state
becomes $9.$ From state $i\geq1$ the system moves to state $i-1$
in the next slot. 

At state $0$, control $g_{0}$ results in ``inactive'' channels.
If control $g_{2}$ is taken, the result is either ``unsuccessful''
or ``successful'' transmission on channel $\ell_{2}$ --- a random
event --- and if control $g_{1}$ is taken, the result is ``successful
transmission'' on channel $\ell_{1}$. Here, a ``successful''
transmission should be taken to mean that a packet will be successfully
delivered to node $b$ if transmitted in the slot (in other words,
a ``good'' underlying transmission link); it does not
preclude the respective control to include a possible action that
does not make a transmission in the slot at all.

The controls under which one of the links is activated are associated
with two actions: a) the action of transmitting a packet on the corresponding
link, if the queue is nonempty and b) the action of not transmitting
a packet (``null'' action). For the control that does
not activate any link,
the associated action set is only the ``null'' action. 

During system operation, there will be a number of packets at the
queue of node $a$. The number of packets in the queue at time $t$
is part of $\mathcal{H}_{t}$, not part of the system state. Based
on $\mathcal{H}_{t}$ and $s_{t}$, a policy takes control $g\in\mathcal{G}_{s_{t}}$
and then an action $\alpha\in\mathcal{A}_{g}$. Depending on the result
of the control, a transmitted packet (if any) may be successfully received,
or erased. 
\end{example}

\textbf{Departures. }There are well-defined times when each arriving
packet is considered to depart from the system. For example, in a
store-and-forward communication network where a packet arrives at
node $i$ and must be delivered to a single node $j,$ it is natural
to consider the departure time as the time at which this packet is
delivered to node $j$. Similarly, if the packet must be multicast
to a subset $\mathcal{K}$ of the nodes, the departure time of the
packet can be defined as the first time at which all nodes in $\mathcal{K}$
receive the packet. However, in some systems several definitions of
departure times may make sense, and the particular choice depends
on the performance measures of interest. As an example, consider the
case where network coding is used to transmit encoded packets. In
this case, a packet $p$ arriving at a single-destination node $j$
may be considered as departed when the
destination node $j$ can decode the packet based on the packets already
received by that node. On the other hand, if the decoded packet is
still needed for decoding of other packets, it may be of interest
to define the departure time of $p$ as the first time the packet
is not needed for further decoding. At any time between the arrival
and departure times of a packet $p,$ we say that $p$ is ``in the
system''.

There may be several restrictions on the policies in $\Pi.$ We assume
that all policies in $\Pi$ have the following features.

\textbf{Features of Admissible Policies}
\begin{enumerate}
\renewcommand{\theenumi}{F\arabic{enumi}}
\item \label{Feature1}At time $t,$ the history of the system up to $t$,
$\mathcal{H}_{t}$ is fully known. 
\item \label{Feature2}At any time $t$ at which there are packets only
at the inputs of the system, it is permissible to take controls and
actions taking into account only the packet at the inputs at time
$t,$ and to proceed without taking into account the rest of history
$\mathcal{H}_{t}$. Formally, for any time $t$ in which the internal (non-input)
queues, if any, are empty, the set of controls and actions available
to a policy may only depend on the current queue state and may not depend
further on $\mathcal{H}_t$.
\item \label{Feature3}If at time $t$ there are $\boldsymbol{k}$ packets
at the inputs of the system, it is permissible to pick any $\boldsymbol{m}\leq\boldsymbol{k}$
packets and continue processing the $\boldsymbol{m}$ packets, along
with other packets that may be in the system, \emph{without taking
into account }the remaining $\boldsymbol{k}-\boldsymbol{m}$
packets. Formally, the set of controls (and actions) available to a policy
must be a superset of the set of controls (and actions, respectively)
that would be available if $\boldsymbol{k}-\boldsymbol{m}$ packets were removed
from the input queues altogether, for any $\boldsymbol{m}\leq\boldsymbol{k}$.
\end{enumerate}

Features \ref{Feature2} and \ref{Feature3} may be natural for many
systems, however, there are systems where they may not be available
to the policies, as the following example shows. 
\begin{example}
\emph{\label{Two-transmitter-Aloha}Two-transmitter Aloha-type system.
}Consider a system consisting of two transmitters attempting to transmit
arriving packets to a single destination. Each transmitter has its
own queue. Activation of both transmitters in the same slot results
in loss of any packet that may be transmitted. We can model this system
by considering that it has a single state, and that the control set
consists of pairs $\left(g_{1},g_{2}\right)$ where $g_{i}=1$
($g_{i}=0)$ indicates that transmitter $i\in\{1,2\}$ becomes active (inactive). 

Consider the following classes of policies, $\Pi_{1},\:\Pi_{2}$:
admissible policies $\pi$ of both classes have Feature \ref{Feature1}.
Also, if only one transmitter queue, say transmitter $1$ queue, has
packets at time $t$, only the transmitter of this queue becomes active,
that is the control $(1,0)$ is chosen. However, the policies in the
two classes differ when both transmitter queues are nonempty. In this
case, policies in $\Pi_{1}$ are free to activate any of the transmitters.
Under policies in $\Pi_{2}$ on the other hand, the controls are chosen
randomly, so that each transmitter becomes active with probability
$q_{t},$ $0\leq q_{t}\leq1$ (and inactive with probability $1-q_{t}$),
$q_{t}$ being the \emph{same }for both transmitters. An action here consists of sending a packet if a transmitter is active.

The policies in $\Pi_{1}$ have Feature \ref{Feature3}, while the
policies in $\Pi_{2}$ do not, since if, e.g. $\boldsymbol{k}=(1,1),$
and control $(1,1)$ is selected, packets from both queues must be
transmitted at the same time, i.e, a policy is not allowed to transmit
first the vector $(1,0)$ and next the vector $(0,1)$. Also, note
that in both cases the policies trivially have feature \ref{Feature2}.

Consider now a third class of policies, $\Pi_{3},$ where policies
act as policies in $\Pi_{2},$ with the following difference: a policy
$\pi\in\Pi_{3}$ selects again a common packet transmission probability
$q$ when both queues are nonempty; however, after a given number $k$ of times this probability has been selected, it must
 thereafter remain fixed and the policy is no longer permitted to change it. For
this class of policies, Feature \ref{Feature2} is not satisfied. 
\end{example}
At the beginning of slot $0$ let the system state be $s$ and let
there be $k_{i}\geq0,\: i=1,...,n$ packets at input $i$ and no arrivals
afterwards, i.e., $A_{i}(0)=k_{i},\: A_{i}(t)=0,\: t=2,3,...$. Let
$T_{s}^{\pi}(\boldsymbol{k})\geq0,\:\boldsymbol{k}\neq\boldsymbol{0}$
be the time it takes until all of these packets depart from the system
under policy $\pi.$ We call $T_{s}^{\pi}(\boldsymbol{k})$ the \emph{evacuation
time} under policy $\pi$ when the system starts in state $s$ with
$\boldsymbol{k}$ packets at the inputs, and denote its average value,
$\bar{T}_{s}^{\pi}\left(\boldsymbol{k}\right)=\mathbb{E}\left[T_{s}^{\pi}(\boldsymbol{k})\right],\:\boldsymbol{k}\neq\boldsymbol{0}$.
It will also be convenient to define $\bar{T}_{s}^{\pi}\left(\boldsymbol{0}\right)=1,$ a convention that has the meaning of advancing one slot whenever the system is empty. 

Let 
\begin{equation}
\bar{T}_{s}^{*}(\boldsymbol{k})=\inf_{\pi\in\Pi}\bar{T}_{s}^{\pi}(\boldsymbol{k})\label{eq:4-1}
\end{equation}
and 
\[
\bar{T}^{*}(\boldsymbol{k})=\max_{s\in\mathcal{S}}\bar{T}_{s}^{*}(\boldsymbol{k}).
\]

We call $\bar{T}^{*}(\boldsymbol{k})$ the \emph{critical evacuation time
function}. It will be seen that under certain statistical assumptions,
this function determines the stability region of the policies under
consideration. 

Note that according to the definition of $\bar{T}_{s}^{*}(\boldsymbol{k})$,
for any $\epsilon>0$ we can always find a policy $\pi$ such that
\begin{equation}
\bar{T}_{s}^{\pi}(\boldsymbol{k})\leq\bar{T}_{s}^{*}(\boldsymbol{k})+\epsilon.\label{eq:3-ff}
\end{equation}
This fact will be used repeatedly in the development that follows. 

Next, we present statistical assumptions regarding the system under
consideration.

\textbf{Statistical Assumptions}
\begin{enumerate}
\renewcommand{\theenumi}{SA\arabic{enumi}}
\item \label{StatAssumption1.0.1} For all $\boldsymbol{k}$
\[
\bar{T}_{s}^{*}(\boldsymbol{k})\leq\infty.
\]

\item \label{StatAssumption1.0}System and arrival process statistics are
known to a policy. 
\item \label{StatAssumption1}Markings (such as service times, permissible
routing paths, etc) associated
with packets arriving to a given input are independent and statistically
identical. Markings across inputs are independent.
\item \label{StatAssumption2}If at the beginning of a slot $t$ the system
state is $s_{t}\in\mathcal{S}$ and control $g_{t}\in\mathcal{G}_{s_{t}}$
is taken, the results at time $t+1$ are independent of the system
history before $t$. However, the system state $s_{t+1}$ and the
results at time $t+1$ may depend on both $s_{t}$ and $g_{t}$. Hence
the system states may be affected by the controls (but not
actions) taken by a policy.
Formally, if $W_{t}$ is the (random) outcome at the end of a slot,
we have for all $t$, 
\[
\Pr\left(W_{t+1},S_{t+1}\left|s_{t},g_{t},\mathcal{H}_{t}\right.\right)=\Pr\left(W_{t+1},S_{t+1}\left|s_{t},g_{t}\right.\right).
\]

\item \label{StatAssumption4}At time $t=0,1,2,\dots$ let there be $\boldsymbol{k}$
packets in the system (where $k_i$ is the number of packets still in the system
that originally arrived at input $i$; they may or may not still be at the input queues). There is a policy
$\pi_{h}$ which can process all these packets until they all depart
from the system by time $t+F^{\pi_{h}}\left(\boldsymbol{k}\right)$
($F^{\pi_{h}}\left(\boldsymbol{k}\right)$ may be random), such that
\begin{equation}
\mathbb{E}\left[F^{\pi_{h}}\left(\boldsymbol{k}\right)\right]\leq C_{1}\sum_{i=1}^{n}k_{i}+C_{0},\label{eq:2-1}
\end{equation}
where $C_{1},\: C_{0}$ are finite constants (which may depend on system
statistics but not on $\boldsymbol{k}$).
\item \label{StatAssumption5}Let $\boldsymbol{e}_{i}$ be the unit $n$-dimensional
vector with $1$ at the $i$-th coordinate and $0$ elsewhere. It holds for all $i=1,..,n,$ and all $\boldsymbol{k}$ and $s$,
\begin{equation}
\bar{T}_{s}^{*}\left(\boldsymbol{k}\right)-\bar{T}_{s}^{*}\left(\boldsymbol{k}+\boldsymbol{e}_{i}\right)\leq D_{0}<\infty.\label{eq:5-3}
\end{equation}

\end{enumerate}

Statistical Assumption \ref{StatAssumption4} is easy to verify in
several systems. For example, in a communication network a policy
that usually satisfies this assumption is the one that picks one of
the $\boldsymbol{k}$ packets, transmits it to its destination, then
picks another packet and so on, until all the packets are delivered
to their destinations. Note that assumption \ref{StatAssumption4}
implies \ref{StatAssumption1.0.1}; we keep assumption \ref{StatAssumption1.0.1}
separate because, as will be seen shortly, only this assumption is needed
to establish the key property (namely, subadditivity) of $\bar{T}^{*}\left(\boldsymbol{k}\right).$

Statistical Assumption \ref{StatAssumption5} is needed to justify
a technical condition in the development that follows. This assumption
may also be easy to verify for several systems. It says
that, if the number of packets at the system inputs at time $0$ is
\emph{increased} by one, then the minimal average evacuation time under
any initial state cannot be \emph{decreased} by more than a fixed amount.
For example, this assumption is always satisfied if $\bar{T}_{s}^{*}\left(\boldsymbol{k}\right)$
is non-decreasing in $\boldsymbol{k}$, i.e., 
\begin{equation}
\bar{T}_{s}^{*}\left(\boldsymbol{k}\right)\leq\bar{T}_{s}^{*}\left(\boldsymbol{k}+\boldsymbol{e}_{i}\right).\label{eq:5-1}
\end{equation}
In particular, it can be easily shown that condition~\eqref{eq:5-1} holds if policies
have the ability to generate ``dummy'' packets (i.e. packets that
bear no information and are used just for policy implementation) during
their operation, a feature that is available in many communication
networks. Indeed, assume that at time $t=0$ the system is in state
$s$ and there are $\boldsymbol{k}$ packets at the system inputs.
Pick $\epsilon>0$ and a policy $\pi$ such that 
\[
\bar{T}_{s}^{\pi}\left(\boldsymbol{k}+\boldsymbol{e}_{i}\right)\leq\bar{T}_{s}^{*}\left(\boldsymbol{k}+\boldsymbol{e}_{i}\right)+\epsilon.
\]
Consider the following policy $\pi_{0}$ for evacuating $\boldsymbol{k}$
packets: generate a ``dummy'' packet for input $i$, place the $\boldsymbol{k}+\boldsymbol{e}_{i}$
packets at the inputs and use policy $\pi$ to evacuate the system.
By construction, $T_{s}^{\pi_{0}}\left(\boldsymbol{k}\right)\leq T_{s}^{\pi}\left(\boldsymbol{k}+\boldsymbol{e}_{i}\right)$
(the inequality may be strict if the departure time of the dummy packet
turns out to be strictly larger that the departure times of the rest
of the packets). Hence, 
\begin{align*}
\bar{T}_{s}^{*}\left(\boldsymbol{k}\right) & \leq\bar{T}_{s}^{\pi_{0}}\left(\boldsymbol{k}\right)\\
 & \leq\bar{T}_{s}^{\pi}\left(\boldsymbol{k}+\boldsymbol{e}_{i}\right)\\
 & \leq\bar{T}_{s}^{*}\left(\boldsymbol{k}+\boldsymbol{e}_{i}\right)+\epsilon.
\end{align*}
Since $\epsilon$ is arbitrary, \eqref{eq:5-1} follows.

To conclude the discussion of Assumption \ref{StatAssumption5}, we provide
an example for which Assumption \ref{StatAssumption5} holds, even though $\bar{T}_{s}^{*}\left(\boldsymbol{k}\right)$
may decrease as $\boldsymbol{k}$ increases. This example is inspired
by~\cite{BacFoss}.
\begin{example}
\label{Example4}Consider a system with two inputs. If packets from
both inputs are processed simultaneously, then both depart after $a$
slots. If a single packet from any of the inputs is processed, then
this packet departs in $A$ slots, where $A>a.$ Admissible policies
may select to transmit pairs of packets (one from each queue) or single
packets. It is easily seen that 
\[
\bar{T}^{*}\left(k_{1},k_{2}\right)=a\min\left\{ k_{1},k_{2}\right\} +A\left|k_{1}-k_{2}\right|.
\]
 Hence, for any $k$, $\bar{T}^{*}\left(k,k+1\right)=ak+A$ and $\bar{T}^{*}\left(k+1,k+1\right)=a\left(k+1\right)<\bar{T}^{*}\left(k,k+1\right).$
On the other hand, we always have, 
\begin{align*}
\bar{T}^{*}\left(k_{1},k_{2}\right)-\bar{T}^{*}\left(k_{1}+1,k_{2}\right) & =a\min\left\{ k_{1},k_{2}\right\} +A\left|k_{1}-k_{2}\right|-a\min\left\{ k_{1}+1,k_{2}\right\} -A\left|k_{1}+1-k_{2}\right|\\
 & \leq A.
\end{align*}

\end{example}

\section{Properties of Critical Evacuation Time Function}

The following property of the critical evacuation time function will play
a key role in the following. 
\begin{lem}
\label{lem:0}The Critical Evacuation Time Function is subadditive,
i.e., the following holds for $\boldsymbol{m}\geq\boldsymbol{0}$,
$\boldsymbol{k}\geq\boldsymbol{0}$ 

\begin{equation}
\bar{T}^{*}(\boldsymbol{k}+\boldsymbol{m})\leq\bar{T}^{*}(\boldsymbol{k})+\bar{T}^{*}(\boldsymbol{m})\label{eq:1}
\end{equation}
\end{lem}
\begin{IEEEproof}
Let $\epsilon>0$ and let the system be in state $s$ at time $0$.
An admissible policy $\pi$ that evacuates $\boldsymbol{k}+\boldsymbol{m}$
packets is the following.

a) Pick an admissible policy $\pi_{\boldsymbol{k}}$ such that, 
\begin{align*}
\bar{T}_{s}^{\pi_{\boldsymbol{k}}}(\boldsymbol{k}) & \leq\bar{T}_{s}^{*}(\boldsymbol{k})+\epsilon/2.
\end{align*}

b) Evacuate the $\boldsymbol{k}$ packets following policy $\pi_{\boldsymbol{k}}$.
According to Feature \ref{Feature3} this is permissible. From Statistical
Assumption \ref{StatAssumption2} we conclude that the average evacuation
time in this case is $\bar{T}_{s}^{\pi_{\boldsymbol{k}}}(\boldsymbol{k})$
. Let $s_{1}$ be the state of the system by time $T_{s}^{\pi_{\boldsymbol{k}}}(\boldsymbol{k}).$
Both $s_{1}$ and $T_{s}^{\pi_{\boldsymbol{k}}}(\boldsymbol{k})$
are known to $\pi_{\boldsymbol{k}}$ (hence to $\pi$), due to Feature
\ref{Feature1}. Note that $s_{1}$ is a random variable that depends on $s$.

c) Again, pick an admissible policy $\pi_{\boldsymbol{m}}$ such that,
\[
\bar{T}_{s_{1}}^{\pi_{\boldsymbol{m}}}(\boldsymbol{m})\leq\bar{T}_{s_{1}}^{*}(\boldsymbol{m})+\epsilon/2,
\]
According to Feature \ref{Feature2}, this choice of $\pi_{\boldsymbol{m}}$
is permissible.

d) Evacuate the $\boldsymbol{m}$ packets following policy $\pi_{\boldsymbol{m}.}$
Due to Statistical Assumption \ref{StatAssumption1} and \ref{StatAssumption2}, the average
evacuation time (given $s_1$) in this case is $\bar{T}_{s_{1}}^{\pi_{\boldsymbol{m}}}(\boldsymbol{m}).$

The average evacuation time of $\pi$ is 
\begin{align}
\bar{T}_{s}^{\pi}\left(\boldsymbol{k}+\boldsymbol{m}\right) & =\bar{T}_{s}^{\pi_{\boldsymbol{k}}}(\boldsymbol{k})+\mathbb{E}\left[\bar{T}_{s_{1}}^{\pi_{\boldsymbol{m}}}(\boldsymbol{m})\right]\nonumber \\
 & \leq\bar{T}_{s}^{*}(\boldsymbol{k})+\mathbb{E}\left[\bar{T}_{s_{1}}^{*}(\boldsymbol{m})\right]+\epsilon,\label{eq:6-1}
\end{align}
where the expectation in (\ref{eq:6-1}) is with respect to random variable $s_1$. Hence,
\begin{align*}
\bar{T}^{*}(\boldsymbol{k}+\boldsymbol{m}) & =\max_{s\in\mathcal{S}}\bar{T}_{s}^{*}(\boldsymbol{k}+\boldsymbol{m})\\
 & \leq\max_{s\in\mathcal{S}}\bar{T}_{s}^{\pi}\left(\boldsymbol{k}+\boldsymbol{m}\right){\rm \:\:\: according}\:{\rm to\:(}\ref{eq:4-1})\\
 & \leq\max_{s\in\mathcal{S}}\left\{ \bar{T}_{s}^{*}(\boldsymbol{k})+\mathbb{E}\left[\bar{T}_{s_{1}}^{*}(\boldsymbol{m})\right]\right\} +\epsilon{\rm \:\:\: according}\:{\rm to\:(}\ref{eq:6-1})\\
 & \leq\max_{s\in\mathcal{S}}\bar{T}_{s}^{*}(\boldsymbol{k})+\max_{s\in\mathcal{S}}\bar{T}_{s}^{*}(\boldsymbol{m})+\epsilon\:\:\:{\rm since}\: \bar{T}_{s_1} \leq \max_{s\in\mathcal{S}}\bar T_s
\end{align*}
Since $\epsilon$ is arbitrary, the lemma follows. 
\end{IEEEproof}
Let $\mathbb{N}_{0}$ and $\mathbb{R}_{0}$ be respectively the set
of nonnegative integers and nonnegative real numbers. We extend the
domain of definition of $\bar{T}^{*}\left(\boldsymbol{k}\right)$
from $\mathbb{N}_{0}^{n}$ to $\mathbb{R}_{0}^{n}$ as follows. For
$\boldsymbol{r}\in\mathbb{R}_{0}^{n},$ let
\begin{equation}
\bar{T}^{*}\left(\boldsymbol{r}\right)=\bar{T}^{*}\left(\left\lceil \boldsymbol{r}\right\rceil \right).\label{eq:8-f}
\end{equation}
The function $\bar{T}^{*}\left(\boldsymbol{r}\right)$ is not necessarily
subadditive in $\mathbb{R}_{0}^{n}$, since, in general, subadditivity at
integer points does not imply subadditivity over $\mathbb{R}_{0}$. For example, the function $f\left(2l\right)=al$
and $f\left(2l+1\right)=al+A$, $l=0,1,...,$ with $a<A,$ is subadditive
in $\mathbb{N}_{0}$, while for $r_{1}=r_{2}=1.5$, $f\left(\left\lceil r_{1}+r_{2}\right\rceil \right)=f(3)=a+A$
and $f\left(\left\lceil r_{1}\right\rceil \right)+f\left(\left\lceil r_{2}\right\rceil \right)=2a<f\left(\left\lceil r_{1}+r_{2}\right\rceil \right).$
However, as the next Lemma shows, $\bar{T}^{*}\left(\boldsymbol{r}\right)$
possesses the basic property of subadditive functions, namely the
asymptotically linear rate of growth. 
\begin{thm}
\label{lem:4-1}For any $\boldsymbol{r}\in\mathbb{R}_{0}^{n}$, the
limit function 
\begin{equation}
\hat{T}(\boldsymbol{r})=\lim_{t\rightarrow\infty}\frac{\bar{T}^{*}\left(t\boldsymbol{r}\right)}{t},\label{eq:3-2}
\end{equation}
exists and is finite, positively homogeneous, convex and Lipschitz
continuous, i.e., it holds 
\[
\left|\hat{T}\left(\boldsymbol{r}\right)-\hat{T}\left(\boldsymbol{s}\right)\right|\leq D\sum_{i=1}^{n}\left|r_{i}-s_{i}\right|,
\]
where $D$ is a positive constant.
 Moreover, for any sequence $\boldsymbol{r}_{t}\in\mathbb{R}_{0}^{n}$
such that
\[
\lim_{t\rightarrow\infty}\boldsymbol{r}_{t}=\boldsymbol{\lambda}<\boldsymbol{\infty,}
\]
it holds
\begin{equation}
\lim_{t\rightarrow\infty}\frac{\bar{T}^{*}\left(t\boldsymbol{r}_{t}\right)}{t}=\hat{T}\left(\boldsymbol{\lambda}\right).\label{eq:10-f}
\end{equation}

\end{thm}
Here, ``positively homogeneous'' means that for any $\rho\geq0,$
\begin{equation}
\hat{T}\left(\rho\boldsymbol{r}\right)=\rho\hat{T}\left(\boldsymbol{r}\right).\label{eq:5-2}
\end{equation}

The proof of Theorem \ref{lem:4-1} is given in the Appendix.

\section{Stability - Necessity\label{sec:Stability---Necessity}}

Let $D_{s,i}^{\pi}(t),\: t\geq1,$ be the number of packet arrivals
at input $i$ that have departed from the system during time slot $t$ under policy $\pi\in\Pi$ when the system starts in state
$s$. Define also $D_{s,i}^{\pi}(0)=0.$ In the following we will
use the notation 
\[
\tilde{A}_{i}\left(t\right)=\sum_{\tau=0}^{t}A_{i}\left(\tau\right),\quad\tilde{D}_{s,i}^{\pi}\left(t\right)=\sum_{\tau=0}^{t}D_{s,i}^{\pi}\left(\tau\right),
\]
to denote the cumulative number or arrivals and departures respectively
up to time $t.$ Hence the number of packet arrivals at input $i$
that are still in the system at time $t$ is $Q_{s,i}^{\pi}\left(t\right)=\tilde{A}_{i}\left(t\right)-\tilde{D}_{s,i}^{\pi}\left(t\right)$
(these packets may
at time $t$ be scattered among internal system queues as well as the original
input queue).
We define
the vector $\boldsymbol{Q}_{s}^{\pi}\left(t\right)=\left(Q_{s,i}^{\pi}\left(t\right)\right)_{i=1}^{n}$ and
the total system occupancy
\[
Q_{s}^{\pi}\left(t\right)=\sum_{i=1}^{n}Q_{s,i}^{\pi}\left(t\right).
\]

Let $\mathcal{M}$ be a probability measure over the space of permissible
arrival processes; in other words, $\mathcal{M}$ captures the statistical
assumptions about the arrival processes, such as the distribution of the
arrival sizes, whether or not the arrivals are independent over time and
between queues, etc. Let $\mathcal{M}_{\boldsymbol{\lambda}}$ be a probability
measure over arrival processes that satisfy ergodicity conditions
\mbox{\eqref{eq:1.0}--\eqref{eq:-1}} with a rate vector $\boldsymbol{\lambda}$.
\begin{defn}
\textbf{\emph{\label{Def:Stab2}System Stability. }}
A policy $\pi\in\Pi$ is called stable
for an arrival rate vector $\boldsymbol{\lambda}\geq\boldsymbol{0,}$
if under any initial system state $s$,
the following holds:
\begin{equation}
\lim_{q\rightarrow\infty}\limsup_{t\rightarrow\infty}\Pr\left(Q_{s}^{\pi}\left(t\right)>q\right)=0\label{eq:6-2}
\end{equation}
(where the probability in~\eqref{eq:6-2} is taken with respect to the arrival
process statistics $\mathcal{M}_{\boldsymbol{\lambda}}$, as well as the 
system internal state transitions).

\end{defn}
The stability region $\mathcal{R}^{\pi}$ of a policy $\pi$ (under
$\mathcal{M}$) is the closure of the set of the arrival rate vectors for
which the policy is stable. The stability region $\mathcal{R}$ of
the system is the closure of the union of $\mathcal{R}^{\pi}, \pi\in\Pi$. 
\footnote{We emphasize that the stability region of a policy
may in general depend on the permitted statistical assumptions about the
arrival processes; for example, a policy may be unstable for a certain rate
vector $\boldsymbol{\lambda}$ if general stationary arrival processes are
allowed, but become stable if the individual queue arrivals are required to be
independent. The above definition of stability is generic and
captures a number of common definitions of stability in the literature, and
the subsequent discussion in this section is orthogonal to any specific
assumptions imposed on the arrival process, beyond the basic ergodicity
condition of
\mbox{\eqref{eq:1.0}--\eqref{eq:-1}}.}

We show in Theorem \ref{lem:3} below that under (\ref{eq:1.0}) and (\ref{eq:-1}),
it holds $\mathcal{R}\subseteq\left\{ \boldsymbol{r}\geq\boldsymbol{0}:\:\hat{T}\left(\boldsymbol{r}\right)\leq1\right\} .$
Furthermore, in section \ref{sub:Epoch-Based-Policy} we show that
under the assumption that the packet arrival vectors are i.i.d. over time,
we also have $\left\{ \boldsymbol{r}\geq0:\:\hat{T}\left(\boldsymbol{r}\right)\leq1\right\} \subseteq\mathcal{R},$
hence, $\mathcal{R}=\left\{ \boldsymbol{r}\geq0:\:\hat{T}\left(\boldsymbol{r}\right)\leq1\right\}$,
and we show an explicit policy called ``Epoch-based'' that is stabilizing.

For the proof of Theorem \ref{lem:3} we need the following lemma.
\begin{lem}
\label{lem:weak->average}If (\ref{eq:6-2}), (\ref{eq:1.0}), (\ref{eq:-1})
hold, then 
\begin{equation}
\lim_{t\rightarrow\infty}\frac{\mathbb{E}\left[Q_{s}^{\pi}\left(t\right)\right]}{t}=0.\label{eq:15new}
\end{equation}
\end{lem}
\begin{IEEEproof}
It follows from (\ref{eq:1.0}), (\ref{eq:-1}) and the corollary
to Theorem 16.14 in \cite{Bi95} that the sequences $\left\{ \tilde{A}_{i}\left(t\right)/t\right\} \: i=1,..,n$
are uniformly integrable, hence the sequence $\left\{ \sum_{i=1}^{n}\tilde{A}_{i}\left(t\right)/t\right\} $
is also uniformly integrable. Since
\[
0\leq\frac{Q_{s}^{\pi}\left(t\right)}{t}\leq\frac{\sum_{i=1}^{n}\tilde{A}_{i}\left(t\right)}{t},
\]
 we conclude that the sequence $\left\{ Q_{s}^{\pi}\left(t\right)/t\right\} $
is also uniformly integrable. We will show in the next paragraph that
$\left\{ Q_{s}^{\pi}\left(t\right)/t\right\} $ converges in probability
to $0.$ Equality (\ref{eq:15new}) will then follow from Theorem
25.12 in \cite{Bi95}.

Pick any $\epsilon>0$ (arbitrarily small) and a $q\geq0$ large enough
so that according to (\ref{eq:6-2}) it holds,
\[
\limsup_{t\rightarrow\infty}\Pr\left\{ Q_{s}^{\pi}\left(t\right)>q\right\} \leq\epsilon
\]
 Since we can pick $t_{0}$ large enough so that $\epsilon t>q,$
$t\geq t_{0},$ we have 
\begin{align*}
\limsup_{t\rightarrow\infty}\Pr\left\{ \frac{Q_{s}^{\pi}\left(t\right)}{t}>\epsilon\right\}  & \leq\limsup_{t\rightarrow\infty}\Pr\left\{ Q_{s}^{\pi}\left(t\right)>q\right\} \\
 & \leq\epsilon
\end{align*}
i.e., $\left\{ Q_{s}^{\pi}\left(t\right)/t\right\} $ converges in
probability to $0.$\end{IEEEproof}
\begin{thm}
\label{lem:3} Let (\ref{eq:1.0}), (\ref{eq:-1}) hold. If $\boldsymbol{r}\in\mathcal{R}$
then, 
\[
\hat{T}\left(\boldsymbol{r}\right)\leq1.
\]
\end{thm}
\begin{IEEEproof}
Pick $\boldsymbol{r}\in\mathcal{R}$. Since $\boldsymbol{r}$ belongs
to the closure of the rates for which the system is stabilizable,
for any $\delta>0$ we can find a $\boldsymbol{\lambda}\geq\boldsymbol{0,}$
$\Vert\boldsymbol{\lambda}-\boldsymbol{r}\Vert\leq\delta,$ for which
the system is stable under some policy $\pi_{0}\in\Pi.$ By continuity
of $\hat{T}\left(\boldsymbol{r}\right)$ it suffices to show that
for any such $\boldsymbol{\lambda},$ 
\begin{equation}
\hat{T}\left(\boldsymbol{\lambda}\right)\leq1.\label{eq:21}
\end{equation}
Let the initial system state be $s\in\mathcal{S}$. Fix an arbitrary
time index $t$ and generate random number of packets $\boldsymbol{A}(0),....,\boldsymbol{A}(t)$
according to the distribution of the arrival processes. Consider that
all $\tilde{\boldsymbol{A}}\left(t\right)=\sum_{\tau=0}^{t}\boldsymbol{A}\left(t\right)$
packets are in the system at the beginning of time and construct the
following evacuation policy $\pi.$

1. Mimic the actions of policy $\pi_{0}$ for up to $t$ time slots,
assuming that the packet arrival process at time $\tau$ is $\boldsymbol{A}\left(\tau\right),\:\tau=1,...,t$.
Due to Statistical Assumption \ref{StatAssumption1.0} and Features
\ref{Feature1}, \ref{Feature3} this mimicking is permissible.\footnote{We
remark that the theorem continues to hold even if \emph{anticipative} policies
are allowed, i.e., if Feature \ref{Feature1} is revised so that the information
available to a policy includes not just the past history up to time $t$, but
future packet arrivals as well.
If $\pi_{0}$ is anticipative, one can accordingly generate random variables
$\boldsymbol{A}\left(\tau\right),\:\tau=t+1,\dots$ so that $\pi$
can mimic $\pi_{0}$ taking into account the future arrivals; the rest of
the proof then remains unchanged.}

2. If all $\tilde{\boldsymbol{A}}\left(t\right)$ packets are transmitted
by time $t$ then the evacuation time of $\pi$ is at most $t.$ Else,
after $t$ time slots there will be $\boldsymbol{Q}_{s}^{\pi_{0}}\left(t\right)>0$
packets in the system. According to Statistical Assumption \ref{StatAssumption4},
pick a policy $\pi_{h}$ to evacuate the $\boldsymbol{Q}_{s}^{\pi_{0}}\left(t\right)$
packets in $F^{\pi_{h}}\left(\boldsymbol{Q}_{s}^{\pi_{0}}\left(t\right)\right)$
slots, where 
\begin{align}
\mathbb{E}\left[F^{\pi_{h}}\left(\boldsymbol{Q}_{s}^{\pi_{0}}\left(t\right)\right)\left|\tilde{\boldsymbol{A}}\left(t\right),\tilde{\boldsymbol{D}}_{s}^{\pi_{0}}\left(t\right)\right.\right] & \leq C_{1}Q_{s}^{\pi_{0}}\left(t\right)+C_{0},~~~~\:{\rm by\:(\ref{eq:2-1})}.\label{eq:17new-1}
\end{align}

The evacuation time of $\pi$ given $\tilde{\boldsymbol{A}}\left(t\right)$ is at
most $t+F^{\pi_{h}}\left(\boldsymbol{Q}_{s}^{\pi_{0}}\left(t\right)\right)$~---
``at most'', because all $\tilde{\boldsymbol{A}}\left(t\right)$
packets may have left before time $t$~--- and hence, taking the conditional
average, we have
\begin{align*}
\bar{T}_{s}^{*}\left(\tilde{\boldsymbol{A}}\left(t\right)\right) & \leq t+\mathbb{E}\left[F^{\pi_{h}}\left(\boldsymbol{Q}_{s}^{\pi_{0}}\left(t\right)\right)\left|\tilde{\boldsymbol{A}}\left(t\right)\right.\right]\\
 & =t+\mathbb{E}\left[\mathbb{E}\left[F^{\pi_{h}}\left(\boldsymbol{Q}_{s}^{\pi_{0}}\left(t\right)\right)\left|\tilde{\boldsymbol{A}}\left(t\right),\tilde{\boldsymbol{D}}_{s}^{\pi_{0}}\left(t\right)\right.\right]\left|\tilde{\boldsymbol{A}}\left(t\right)\right.\right]\\
 & \leq t+C_{1}\mathbb{E}\left[Q_{s}^{\pi_{0}}\left(t\right)\left|\tilde{\boldsymbol{A}}\left(t\right)\right.\right]+C_{0}\:{\rm by\:(\ref{eq:17new-1})}
\end{align*}
Next, using the last inequality, 
\begin{align*}
\bar{T}^{*}\left(\tilde{\boldsymbol{A}}\left(t\right)\right) & =\max_{s\in\mathcal{S}}\left\{ \bar{T}_{s}^{*}\left(\tilde{\boldsymbol{A}}\left(t\right)\right)\right\} \\
 & \leq t+C_{1}\max_{s\in\mathcal{S}}\mathbb{E}\left[Q_{s}^{\pi_{0}}\left(t\right)\left|\tilde{\boldsymbol{A}}\left(t\right)\right.\right]+C_{0}\\
 & \leq t+C_{1}\sum_{s\in\mathcal{S}}\mathbb{E}\left[Q_{s}^{\pi_{0}}\left(t\right)\left|\tilde{\boldsymbol{A}}\left(t\right)\right.\right]+C_{0}.
\end{align*}
Taking expectations with respect to $\tilde{\boldsymbol{A}}(t)$ and dividing by $t$, we have
\begin{equation}
\mathbb{E}\left[\frac{\bar{T}^{*}\left(\frac{\tilde{\boldsymbol{A}}\left(t\right)}{t}t\right)}{t}\right]\leq1+C_{1}\sum_{s\in\mathcal{S}}\frac{\mathbb{E}\left[Q_{s}^{\pi_{0}}\left(t\right)\right]}{t}+\frac{C_{0}}{t}\label{eq:17new}
\end{equation}
Since
\[
\lim_{t\rightarrow\infty}\frac{\tilde{\boldsymbol{A}}\left(t\right)}{t}=\boldsymbol{\lambda},\:\mathrm{by\:(\ref{eq:1.0}),}
\]
using (\ref{eq:10-f}) from Theorem \ref{lem:4-1} we then obtain,
\[
\lim_{t\rightarrow\infty}\frac{\bar{T}^{*}\left(\frac{\tilde{\boldsymbol{A}}\left(t\right)}{t}t\right)}{t}=\hat{T}\left(\boldsymbol{\lambda}\right)
\]
Hence, 
\begin{align*}
\hat{T}\left(\boldsymbol{\lambda}\right) & =\mathbb{E}\left[\lim_{t\rightarrow\infty}\frac{\bar{T}^{*}\left(\frac{\tilde{\boldsymbol{A}}\left(t\right)}{t}t\right)}{t}\right]\\
 & \leq\liminf_{t\rightarrow\infty}\mathbb{E}\left[\frac{\bar{T}^{*}\left(\frac{\tilde{\boldsymbol{A}}\left(t\right)}{t}t\right)}{t}\right]\:\:\:\mathrm{by\: Fatou's\; lemma}\\
 & \leq\liminf_{t\rightarrow\infty}\left(1+C_{1}\sum_{s\in\mathcal{S}}\frac{\mathbb{E}\left[Q_{s}^{\pi_{0}}\left(t\right)\right]}{t}+\frac{C_{0}}{t}\right)\:\:\:\mathrm{by\;(\ref{eq:17new})}\\
 & =1+C_{1}\sum_{s\in\mathcal{S}}\lim_{t\rightarrow\infty}\frac{\mathbb{E}\left[Q_{s}^{\pi_{0}}\left(t\right)\right]}{t}+\lim_{t\rightarrow\infty}\frac{C_{0}}{t}\:\mathrm{by\:(\ref{eq:15new})}\\
 & =1\:\mathrm{by\:(\ref{eq:15new})}
\end{align*}

\end{IEEEproof}
We note that there are classes of policies for which the limit $\hat{T}\left(\boldsymbol{\lambda}\right)$
can be formally defined, but Theorem \ref{lem:3} does not hold in
all its generality since some of the Features of admissible policies
in Section \ref{sec:System-Model} are not satisfied. The next example
shows the case where Feature \ref{Feature3} is not satisfied. 
\begin{example}
\label{Example-necessity}Consider the following system. There are
two inputs. Policies may decide to process no packets in a slot, otherwise
processing of packets must obey the following rule. If only one of
the inputs has packets a single packet from the nonempty input is
processed in 1 time slot. If on the other hand both queues are nonempty,
then pairs of packets from both queues must be processed in $3$ time
slots. This system is a simplified version of the system in Example
\ref{Two-transmitter-Aloha} and the specified policies do not satisfy
Feature \ref{Feature3}. It can be easily seen that 
\[
\bar{T}^{*}\left(k_{1},k_{2}\right)=3\min\left\{ k_{1},k_{2}\right\} +\left|k_{1}-k_{2}\right|,
\]
hence formally, 
\[
\hat{T}\left(r_{1},r_{2}\right)=3\min\left\{ r_{1},r_{2}\right\} +\left|r_{1}-r_{2}\right|.
\]
The region $\hat{T}\left(r_{1},r_{2}\right)\leq1$ is described by
\begin{equation}
\left\{ \boldsymbol{r}\geq\boldsymbol{0}:r_{1}+2r_{2}\leq1,\:{\rm and\;}r_{1}\geq r_{2}\right\} \cup\left\{ \boldsymbol{r}\geq\boldsymbol{0}:2r_{1}+r_{2}\leq1,\:{\rm and\;}r_{2}\geq r_{1}\right\} \label{eq:17-ff}
\end{equation}
Clearly, the vector $\left(1/2,1/2\right)$does not belong in this
region. Consider, however that 1 packet arrives in even slots to input
1 and 1 packet in odd slots to input 2, hence the arrival rate vector
is $\left(1/2,1/2\right)$. Then simply processing immediately the
arriving packets results in a stable policy. 

Notice also that the region in (\ref{eq:17-ff}) is not convex, while
the region in Theorem \ref{lem:3} is convex since $\hat{T}\left(r_{1},r_{2}\right)$
is convex.
\end{example}
The arrival processes in the previous example are not stationary,
hence one may wonder whether imposing slightly stronger assumptions
on the arrival processes would render the claim of Theorem \ref{lem:3} valid. An
example is presented below, where the arrival processes are i.i.d.
but Theorem \ref{lem:3} still does not hold since admissible policies do
not satisfy Feature \ref{Feature3}.
\begin{example}
\label{exampleIndependentArrivals}Let $M>1$ and consider a system
with a single input and the following restriction on the policies.
If the number of packets in the inputs is 
\[
k=lM+\upsilon,\:0\leq\upsilon\leq M-1,
\]
then a policy may either decide to idle in a slot or to transmit $m$
packets, $1\leq m\leq M+\upsilon$ in which case it takes $l$ slots
to process all $m$ packets. Under this restriction we have 
\begin{align*}
\bar{T}^{*}\left(k\right) & =\sum_{i=1}^{l}i,\\
 & =\frac{l\left(l+1\right)}{2}
\end{align*}
hence, 
\begin{align*}
\hat{T}\left(r\right) & =\lim_{t\rightarrow\infty}\frac{\bar{T}^{*}\left(\left\lceil tr\right\rceil \right)}{t}\\
 & =\lim_{t\rightarrow\infty}\frac{1}{t}\left(\frac{\left(\left\lceil tr\right\rceil -\upsilon_{t}\right)\left(\left\lceil tr\right\rceil -\upsilon_{t}+1\right)}{2M^{2}}\right)\\
 & =\infty.
\end{align*}
Applying formally Theorem \ref{lem:3} we deduce that the system is
unstable for any positive arrival rate. Consider, however, that the
arrival process is i.i.d but bounded, such that at most $2M-1$ packets
may arrive at the beginning of each slot (including slot 0,
i.e.\ to be in the system when it commences operation). Then the policy that
transmits all the packets immediately is stable, i.e., under the stated
conditions on arrival process statistics, the system is stable for any arrival
rate $\lambda\leq2M-1.$ 
\end{example}
For the systems described in the last two examples, there were rates
outside the region obtained by using formally $\hat{T}\left(\boldsymbol{r}\right),$
for which the systems were stabilizable. The next example shows an opposite
case, namely where the system is unstable for rates inside the formally obtained
region
(again, due to not satisfying Feature \ref{Feature3}).
\begin{example}
\emph{System with priorities and switchover times. }Consider a single
server with two inputs, where arrivals at input 1 have priority over
arrivals at input 2. If there are packets from input 1 in the system,
one of these packets must be served. Packets from input 2 may be delayed
by a policy. Packets are of length 1 slot. There is a preparatory time
of 1 slot to set the system to serve packets from a given input. Hence,
when the system changes from serving packets of one input to serving
packets of the other input, there is an idle slot. The system may
start by having the server ready to serve one of the two inputs. 

The system has two states, $s_{1},\: s_{2},$ where state $s_{i}$
means that the server is set to serve packets of input $i.$ For this
system, we have 
\[
\bar{T}_{s_{1}}^{*}\left(k_{1},k_{2}\right)=\begin{cases}
k_{1}+1+k_{2} & {\rm if\:}k_{2}\ne0\\
k_{1} & {\rm if\:}k_{2}=0
\end{cases}
\]
 and
\[
\bar{T}_{s_{2}}^{*}\left(k_{1},k_{2}\right)=\begin{cases}
1+k_{1}+1+k_{2} & {\rm if\:}k_{1}\neq0,\: k_{2}\ne0\\
1+k_{1} & {\rm if\:}k_{2}=0\\
k_{2} & {\rm if\:}k_{1}=0
\end{cases}
\]
Hence, 
\[
\hat{T}\left(r_{1},r_{2}\right)=r_{1}+r_{2}
\]
 and the region obtained formally is 
\[
\left\{ \boldsymbol{r}\geq\boldsymbol{0}:\: r_{1}+r_{2}\leq1\right\} .
\]
Consider, however an arrival pattern where the system starts at state
$s_{1}$, and a single packet arrives at input 1 at every $t=4k,\: k=0,1,...$;
hence $\lambda_{1}=.25$. Packets at input $2$ arrive according
to an i.i.d process of rate $\lambda_{2}>.5.$ It can be easily checked
that in any interval $[4k,\:4k+8),$ the number of packets served
from input 2 cannot be larger than 4, hence the departure rate for
packet at input $2$ cannot be more than .5 and the system is unstable,
even though $\lambda_{1}+\lambda_{2}<1$.

\end{example}

One may wonder whether if the initial state of the system at time $t=0$ is
fixed, say $s(0)=s_0$, then stability is determined by
$\bar{T}_{s_0}^{*}\left(\boldsymbol{k}\right)$ only. The following final
example illustrates that this is not always the case, i.e.\ the condition of
theorem~\ref{lem:3} applies to the critical (worst-case) evacuation time
function, and not just the evacuation time function corresponding to $s_0$.
\begin{example}
\label{Example:InitStates.}Consider a system with two servers, where
server 1 takes $l$ slots to serve a packet, and server 2 takes $L>l$ slots.
The system can be in one of three states, (0,0), (1,0), (0,1), where
0 denotes an inactive and 1 denotes an active server. Suppose that there are no
(or null) controls, and that state transitions are random with the
following transition probabilities. 
\[
\Pr\left\{ \left(1,0\right)\left|\left(0,0\right)\right.\right\} =\Pr\left\{ \left(0,1\right)\left|\left(0,0\right)\right.\right\} =\Pr\left\{ \left(0,0\right)\left|\left(0,0\right)\right.\right\} =\frac{1}{3},\:\Pr\left\{ \left(1,0\right)\left|\left(1,0\right)\right.\right\} =\Pr\left\{ \left(0,1\right)\left|\left(0,1\right)\right.\right\} =1.
\]
 If the system starts at state $\left(0,0\right)$, it takes on average
$1.5$ slots to move to one of the other states, and the transition to
either state occurs with equal probability. Then, since no further change of
states occurs afterwards, it will take either $lk$ or $Lk$ slots
to evacuate $k$ packets. Hence, 
\[
\bar{T}_{\left(0,0\right)}^{*}\left(k\right)=\frac{3}{2}+\frac{l+L}{2}k.
\]

It can also be easily verified that 
\[
\bar{T}_{\left(1,0\right)}^{*}\left(k\right)=lk
\]
\[
\bar{T}_{\left(0,1\right)}^{*}\left(k\right)=Lk
\]
 hence, 
\[
\bar{T}^{*}\left(k\right)=\max\left\{ \frac{3}{2}+\frac{l+L}{2}k,\: lk,\: Lk\right\} 
\]
 and $\hat{T}\left(r\right)=Lr$, which results in the stability condition,
$\lambda\leq\frac{1}{L}.$ 

Assume now that the system starts in state $s=(0,0)$ (an initial state that may
be ``natural'' in some sense), and formally use
$\bar{T}_{\left(0,0\right)}^{*}\left(k\right)$
in place of $\bar{T}^{*}\left(k\right).$ Then, we would conclude
that $\hat{T}\left(r\right)=\frac{l+L}{2}r$ and hence that the system
is stable when 
\[
\lambda\leq\frac{2}{l+L}.
\]
This, however, is wrong since for $\frac{2}{l+L}>\lambda>\frac{1}{L}$,
under state transition $\left(0,0\right)\rightarrow\left(0,1\right)$,
an event of positive probability, the input rate will be larger than
the output rate. \end{example}

\section{Epoch Based Policy - Sufficiency\label{sub:Epoch-Based-Policy}}

In this section, we consider a specific policy which we henceforth refer to as
an \emph{Epoch-Based} policy. The idea of the policy (which is defined formally
below) is to divide the time into `epochs' and focus on the efficient
evacuation of packets present in the system at the start of an epoch, while
ignoring any new packets that arrive during the epoch. The main result of this
section is that, for the special case of independent and identically
distributed (i.i.d) arrival processes, the epoch-based policy is
throughput-optimal, provided that the underlying evacuation policy within each
epoch is efficient (i.e., informally, minimizes the expected evacuation time
for the packets present at the start of the epoch). More precisely, in this
section we make the assumption that the arrival process vectors
$\boldsymbol{A}(t)$ are i.i.d with respect to time for $t=1,2,\dots$ (for a
given time slot $t$, the components of the vector $\boldsymbol{A}(t)$ may be
dependent; also, the initial number of packets in the system at $t=0$, namely
$\boldsymbol{A}(0)$, can be arbitrary and is not required to have the same
distribution as for $t\ge{1}$). We then show that the epoch-based policy is
stabilizing for any such arrival processes if the arrival rate
$\boldsymbol{\lambda}$ satisfies $\hat{T}\left(\boldsymbol\lambda\right)<1$.

Consider the set 
\[
\mathcal{R}_{l}=\left\{ \boldsymbol{\lambda}\geq\boldsymbol{0}:\:\hat{T}\left(\boldsymbol{\lambda}\right)<1\right\} .
\]
This set is nonempty, since
\begin{equation}
\hat{T}\left(\boldsymbol{0}\right)=\lim_{t\rightarrow\infty}\frac{\bar{T}^{*}\left(t\cdot\boldsymbol{0}\right)}{t}=0,\label{eq:18e}
\end{equation}
 hence $\boldsymbol{0}\in\mathcal{R}_{l}$. We will construct a
policy that is stable for any $\boldsymbol{\lambda}\in \mathcal{R}_{l}$.
The continuity, convexity of $\hat{T}\left(\boldsymbol{\lambda}\right)$
and (\ref{eq:18e}) imply that the closure of $\mathcal{R}_{l}$ is
the set $\left\{ \boldsymbol{\lambda}\geq\boldsymbol{0}:\:\hat{T}\left(\boldsymbol{\lambda}\right)\leq1\right\} $
and hence,
\[
\left\{ \boldsymbol{\lambda}\geq\boldsymbol{0}:\hat{T}\left(\boldsymbol{\lambda}\right)\leq1\right\} \subseteq\mathcal{R}.
\]
 Combined with the necessity result of Section \ref{sec:Stability---Necessity},
we then conclude that 
\[
\mathcal{R}=\left\{ \boldsymbol{\lambda}\geq\boldsymbol{0}:\hat{T}\left(\boldsymbol{\lambda}\right)\leq1\right\} .
\]

We now present a policy that stabilizes the system for any $\boldsymbol{\lambda}\in\mathcal{R}_{l}$,
that is, 
\begin{equation}
\hat{T}\left(\boldsymbol{\lambda}\right)<1.\label{eq:23}
\end{equation}
A version of this policy was used in \cite{Sagduyu09} to
provide a stabilizing policy for a two-user broadcast erasure channel
with feedback.
\begin{defn}
\textbf{\emph{Epoch-Based Policy }}\textbf{$\pi_{\epsilon}$}: Pick
$\epsilon>0$ such that 
\[
0<\epsilon<1-\hat{T}\left(\boldsymbol{\lambda}\right),
\]
and for each $\boldsymbol{k}$ and $s$, pick an evacuation policy
$\pi_{\boldsymbol{k},s}$ such that 
\begin{equation}
\bar{T}_{s}^{\pi_{\boldsymbol{k,s}}}\left(\boldsymbol{k}\right)\leq\bar{T}_{s}^{*}\left(\boldsymbol{k}\right)+\epsilon.\label{eq:ll22}
\end{equation}

Policy $\pi_{\epsilon}$ operates recursively in (random) time intervals
$[t_{m-1},t_{m}),\: m=1,...,$ called ``epochs'', as follows. Epoch
$1$ starts at time $t_{0}=0$ at state $S_{0}=s_{0}$ with $\tilde{\boldsymbol{A}}\left(0\right)=\boldsymbol{A}\left(0\right)=\boldsymbol{k}$
packets at the inputs; policy $\pi_{\boldsymbol{k,s}}$ is used to
evacuate the $\boldsymbol{k}$ packets by time $t_{1}=T_{s}^{\pi_{\boldsymbol{k,s}}}\left(\boldsymbol{k}\right),$
while any new packet arrivals during the epoch are kept at the
inputs, but excluded from processing. Let $S_{m}$ be the state of the system
at time $t_{m}.$
Epoch $m+1,\: m\geq1$ starts at time $t_{m}$ with $\boldsymbol{k}_{m}=\tilde{\boldsymbol{A}}\left(t_{m}\right)-\tilde{\boldsymbol{A}}\left(t_{m-1}\right)$
packets at the inputs and policy $\pi_{\boldsymbol{k}_{m},S_{m}}$
is used to evacuate the $\boldsymbol{k}_{m}$ packets by time $t_{m+1}.$ 
\end{defn}
Let $T_{m}=t_{m}-t_{m-1}, m=1,2,\dots$ be the length of the $m$-th
epoch. Since the arrival process vector is i.i.d, if policies satisfy the Basic Features and the Statistical Assumptions of Section
\ref{sec:System-Model}, the process $\left\{ \left(T_{m},\: S_{m}\right)\right\} _{m=1}^{\infty}$
constitutes a (homogeneous) Markov chain with stationary
transition
probabilities. Note that with this formulation, the initial state of
the Markov chain, $\left(T_{1},S_{1}\right),$ is a random variable
whose distribution depends on $\boldsymbol{A}\left(0\right)$ and
$s_{0}$. 

The main result of this section is the following. 
\begin{thm}
\label{thm:15}For any $\boldsymbol{\lambda}\geq\boldsymbol{0}$ such
that 
\begin{equation}
\hat{T}\left(\boldsymbol{\lambda}\right)<1,\label{eq:31new-1}
\end{equation}
policy $\pi_{\epsilon}$ stabilizes the system. 
\end{thm}
The proof of this theorem is given in the Appendix. 

\textbf{Remarks}
\begin{enumerate}
\item The epoch-based policy is non-anticipative (it does not require knowledge
of future packet arrivals), but is sufficient to attain the stability region
even if anticipative policies are allowed, as explained in the footnote in
the proof of Theorem~\ref{lem:3}. Thus, the ability to anticipate future
packet arrivals is not required for throughput optimality.

\item Note that stability depends solely on the fact that inequality
(\ref{eq:ll22}) holds for large enough $\left|\boldsymbol{k}\right|$. Hence,
for the epoch based policy to be stable, it is sufficient for policies
$\pi_{\boldsymbol{k,s}}$ to satisfy (\ref{eq:ll22}) only for large enough
$\left|\boldsymbol{k}\right|$. In other words, asymptotically optimal
evacuation policies can be used to construct stabilizing epoch based policies.

\item The requirement for the arrival process to be i.i.d. only applies for
$t\ge 1$; the initial queue lengths, namely $\boldsymbol{A}(0)$, may have any
distribution that is not necessarily the same as for $\boldsymbol{A}(t),
t=1,2,\dots$. 
By induction, it is easy to extend
the ``exemption'' up to any finite $t_0$ and only require the arrival process
to be i.i.d. for $t>t_0$.

\item Similarly, the proof can be easily extended to the case where the
arrival process is not i.i.d for individual time slots but is ``block-i.i.d''
with block length $D$; in other words, where the vectors
$\left\langle \boldsymbol{A}(i\cdot{D}+1),\dots,\boldsymbol{A}\left((i+1)\cdot{D}\right)\right\rangle$
are i.i.d with respect to $i$ for $i=1,2,\dots$ (but arrivals may be
interdependent within a ``time block'' $iD+1\le t\le (i+1)D$). 
This is achieved via time scaling by a factor of $D$, namely enforcing epoch
durations to be multiples of $D$ (by simply requiring the epoch-based policy to
wait until the next multiple of $D$ after all packets from the start of the
epoch are evacuated), which allows the Markovian nature of the system to be
maintained.\\
We conjecture that the epoch-based policy can be shown to be stabilizing for
any general stationary and ergodic arrival process, but the necessary extension
of the proof remains open at this stage.

\item A policy, which seems to be more amenable to analysis under stationary and ergodic arrivals is a \emph{frame-based policy} which operates in periods. At each period $n$, beginning at time $S_n$, a number of packets are processed. The packets under processing in period $n$ have all arrived in the system before $S_n$ and correspond to a frame of arrivals of fixed duration $F$. In particular, during the $n^{\text{th}}$ period, only arrivals from the frame $[I_{n-1},I_n)$ are processed, where $I_n-I_{n-1}=F$, hence $I_n\triangleq nF$. The time to evacuate all the arrivals in the interval $[I_{n-1},I_n)$ is random, depending on the number of arrivals as well as other random events and we denote it with $T_n(F)$. 
Note that if there is only one system state, then if the arrival process is stationary, $T_n (F)$ is a stationary process as well.

Before the start of period $n+1$, a waiting time is added if $S_{n}+T_n(F)<I_n+F$. 
This waiting is imposed in order to ensure that $I_{n+1}-I_n=F $.
By letting $D_n=S_n-I_{n}$ denote the lag process, it can be seen that 
\[
D_{n+1}=\left(D_{n}+T_n(F)-F\right)^+.
\]
Note that this equation is of the same form as the recursion relating the queue size in a discrete G/G/1 queue with "arrival rate" (per slot) $T(F)$ and "service rate" $F$. Note that if $T(\lambda) < 1$, then by picking $F$ large enough, we can ensure that $\bar{T}(F) < F$  i.e., "arrival rate" is less than the "service rate". 
We conjecture that this policy stabilizes the queue sizes under stationary and ergodic arrivals. However the policy is unattractive in practice since it induces very large delays even for small arrival rates.
\end{enumerate}

\section{Application: Capacity and Stability regions of Broadcast
Erasure Channel with Feedback }

Consider a communication system consisting of a single transmitter
and a set $\mathcal{N}\stackrel{\vartriangle}{=}\{1,2,\ldots,n\}$
of receivers/users (we hereafter use these two terms interchangeably).
The transmitter has $n$ infinite queues where packets destined to
each of the receivers are stored. Packets consist of $L$ bits and
are transmitted within one slot. The channel is modeled as memoryless
broadcast erasure (BE), so that each broadcast packet is either received
unaltered by a user or is ``erased'' (i.e.~the user does not receive
the packet, but knows that a packet was sent). The latter case is
equivalent to considering that the user receives the special symbol
$E$, which is distinct from any other possible transmitted symbol
and does not map to a physical packet (since it models an erasure).
To complete the description of the system we also need to specify
the outputs when no packet is sent by the transmitter, i.e., the slot
is empty: in this case we assume that all receivers realize that the
slot is empty. An empty slot will be denoted by $\varnothing$. Equivalently,
we may view ``no transmission'' as transmission of a special symbol
$\varnothing.$

In information-theoretic terms, the broadcast erasure channel under
consideration is described by the tuple $(\mathcal{X},(\mathcal{Y}_{i}\in\mathcal{N}),\: p(\boldsymbol{Y}_{l}|X_{l}))$,
where $\mathcal{X}$ is the input symbol alphabet, $\mathcal{Y}_{i}=\mathcal{Y}=\mathcal{X}\cup\{E\}$
is the output symbol alphabet for user $i$, and $p(\boldsymbol{Y}_{l}|X_{l})$
is the probability of having, at slot $l$, output $\boldsymbol{Y}_{l}=(Y_{i,l},i\in\mathcal{N})$
for a broadcast input symbol $X_{l}$. The memoryless property implies
that $p(\boldsymbol{Y}_{l}|X_{l})$ is independent of $l$, so that
it is simply written as $p(\boldsymbol{Y}|X)$. We denote by $\epsilon_{\mathcal{N}_E},\:\mathcal{N}_E\subseteq\mathcal{N},$
the (common) probability that a transmitted packet (i.e. a symbol
in $\mathcal{X}-\left\{ \varnothing\right\} $) is erased by all users in the
set $\mathcal{N}_E$. To avoid unnecessary complications we assume in
the following that $\epsilon_{\{i\}}<1$ for all $i.$ Note that for
the empty slot (symbol $\varnothing$) we have
$\Pr(\boldsymbol{Y}=(\varnothing,...\varnothing)\left|X=\varnothing\right.)=1$.

We assume that there is feedback from the users to the transmitter,
so that at the end of each slot $l$, all users inform the transmitter
whether the symbol was received or not (essentially, a simple ACK/NACK)
through an error-free zero-delay control channel. 

We define two regions for this channel, the information theoretic
capacity region, and the stability region. The information theoretic
capacity region describes transmission rates under which it is possible
to transmit sets of messages (one for each user) placed at the transmitter
by using proper encoding, so that all users receive the messages destined
to them with arbitrarily small probability of error. For the stability
region, Definition \ref{Def:Stab2} is used, under the assumption that packets arrive randomly
to the system. We assume that packets are transmitted using
a proper encoding, such that they are decoded by the receivers with \emph{zero}
probability of error.

We now give a precise definition of the two regions, and show in the following that they are identical.

\subsection*{Information theoretic capacity region }

A channel code, denoted as $c_{l}=(M_{1},\ldots,M_{n},l)$, for the
broadcast channel with feedback is defined as the aggregate of the
following components (this is an extension of the standard capacity
definition of \cite{Cover_info} to $n$ users): 
\begin{itemize}
\item Message sets $\mathcal{W}_{i}$ of size $\left|\mathcal{W}_{i}\right|=M_{i}$
for each user $i\in\mathcal{N}$, where $\left|\cdot\right|$ denotes
set cardinality. Denote the message that needs to be communicated
as $\boldsymbol{W}\stackrel{\vartriangle}{=}(W_{i},i\in\mathcal{N})\in\mathcal{W}$,
where $\mathcal{W}=\mathcal{W}_{1}\times\ldots\times\mathcal{W}_{n}$.
For our purposes it is helpful to interpret the message set $\mathcal{W}_{i}$
as follows: assume that user $i$ needs to decode a given set $\mathcal{K}_{i}$
of $L$-bit packets. Then, $\mathcal{W}_{i}$ is the set of all possible
$\left|\mathcal{K}_{i}\right|L$ bit sequences, so that it holds $\left|\mathcal{W}_{i}\right|=M_{i}=2^{\left|\mathcal{K}_{i}\right|L}$.
Henceforth we will assume this relation. 
\item An encoder that transmits, at slot $t$, a symbol $X_{t}=f_{t}(\boldsymbol{W},\boldsymbol{\hat{Y}}^{t-1})$,
based on the value of $\boldsymbol{W}$ and all previously gathered
feedback $\boldsymbol{\hat{Y}}^{t-1}\stackrel{\vartriangle}{=}(\boldsymbol{Y}_{1},\ldots,\boldsymbol{Y}_{t-1})$,
$\boldsymbol{Y}_{k}=\left(Y_{1,k},...,Y_{n,k}\right)$.
$X_{1}$ is a function of $\boldsymbol{W}$ only. A total of $l$
symbols are transmitted for message $\boldsymbol{W}$.
\item $n$ decoders, one for each user $i\in\mathcal{N}$, represented by
the decoding functions $g_{i}:\mathcal{Y}^{l}\to\mathcal{W}_{i}$
that map $Y_{i}^{l}$, where $Y_{i}^{l}\stackrel{\vartriangle}{=}(Y_{i,1},\ldots,Y_{i,l})$
is the sequence of symbols received by user $i$ during the $l$ slots,
to a message in $\mathcal{W}_{i}$.
\end{itemize}
In the following we write $(M_{1},\ldots,M_{n},l)$ to denote the
code $c_{l}$, with the understanding that the full specification
requires all the components described above. The probability of erroneous
decoding is defined as $q_{l}^{e}=\Pr(\cup_{i\in\mathcal{N}}\{g_{i}(Y_{i}^{l})\neq W_{i}\})$,
where it is assumed that the messages are selected according to the
uniform distribution from $\mathcal{W}$. The rate $\boldsymbol{R}$
for this code, measured in information bits per transmitted symbol,
is now defined as the vector $\boldsymbol{R}=(R_{i}:i\in\mathcal{N})$
with $R_{i}=(\log_{2}M_{i})/l$. Hence, it holds $R_{i}=\left|\mathcal{K}_{i}\right|L/l=r_{i}L$,
where $r_{i}=\left|\mathcal{K}_{i}\right|/l$ is the rate of the code
in packets per slot, and the bits of each packet are uniformly distributed
and independent of the bits of the other packets. For our purposes,
it will be convenient to define the capacity region of the system
in terms of the rate vector $\boldsymbol{r}=\boldsymbol{R}/L.$

A vector rate $\boldsymbol{r}=(r_{1},\ldots,r_{n})$
is achievable if there exists a sequence $\left\{ c_{l}\right\} _{l=1}^{\infty}$
of codes $(2^{\left\lceil lr_{1}\right\rceil L},\ldots,2^{\left\lceil lr_{n}\right\rceil L},l)$
such that $q_{l}^{e}\to0$ as $l\to\infty$. The capacity region $\mathcal{C}$
of the system is the closure of the set of achievable rates.

\subsection*{Stochastic Arrivals: Definitions of admissible policies }

As in Section \ref{sec:System-Model}, we assume that packets arrive
randomly to the system according to the stochastic process $\boldsymbol{A}\left(t\right)$
and are stored in infinite buffers at the transmitter. We denote by
$\boldsymbol{\mathcal{A}}\left(t\right)$ the content of these messages,
i.e., $\boldsymbol{\mathcal{A}}\left(t\right)=\left(\boldsymbol{\mathcal{A}}_{1}\left(t\right),....,\boldsymbol{\mathcal{A}}_{n}\left(t\right)\right)$
where $\boldsymbol{\mathcal{A}}_{i}\left(t\right)=\left(p_{i,1}\left(t\right),...,p_{i,A_{i}(t)}\left(t\right)\right),$
and $p_{i,j}\left(t\right)$ denotes the sequences of bits corresponding
to the $j$th packet with destination node $i$ that arrived at the
transmitter at time $t$~--- if no packets arrive we consider that $\boldsymbol{\mathcal{A}}_{i}\left(t\right)$
is the empty set. We assume that $p_{i,j}\left(t\right)$
are uniformly distributed and mutually independent. 
We denote $\boldsymbol{\hat{\mathcal{A}}}^{t}\triangleq(\boldsymbol{\mathcal{A}}\left(0\right),\dots,\boldsymbol{\mathcal{A}}\left(t\right))$
to be the contents of all packet arrivals up to time $t$.

An admissible policy consists of 
\begin{itemize}
\item An encoder that transmits, at slot $t$, a symbol $X_{t}=f_{t}(\boldsymbol{\hat{Y}}^{t-1},
\boldsymbol{\hat{\mathcal{A}}}^{t})$,
based on all previously gathered feedback, $\boldsymbol{\hat{Y}}^{t-1}=(\boldsymbol{Y}_{1},\ldots,\boldsymbol{Y}_{t-1})$,
and the contents of packet arrivals up to time $t$, $\boldsymbol{\hat{\mathcal{A}}}^{t}$. 
\item $n$ decoders, one for each user $i\in\mathcal{N}$, represented by
the decoding set-valued functions $g_{i,t}\left(Y_{i}^{t}\right)$
that at time $t$ maps $Y_{i}^{t}$ to a subset 
of the packets that have arrived up to time $t-1$ with destination
node $i,$ i.e., 
\[
\mathcal{D}_{t}\subseteq\left\{ p_{i,j}\left(\tau\right):\:\tau\leq t-1,\:1\leq j\leq A_{i}\left(t\right)\right\} .
\]

\end{itemize}
A packet is decoded the first time it is included in $\mathcal{D}_{t}$.
We set the requirement that packet decoding is correct with probability
one. Note that there is at least one policy that satisfies this requirement:
this is the time-sharing policy where packets destined to destination
$i$ are (re)transmitted, in First-Come-First-Served order, until successful
reception in slots specifically assigned to $i$, say slots
$jn+i-1\:,j=0,1,....$.
We call such a policy One By One (OBO) policy, $\pi_{O}$. For this
policy it can be easily seen that 
\[
\bar{T}_{\pi_{O}}\left(\boldsymbol{k}\right)\leq\sum_{i=1}^{n}C_{i}k_{i}+C_{0},
\]
where $C_{i}$ depends on the erasure probabilities, but not on $\boldsymbol{k}.$
Hence, $\pi_{O}$ satisfies (\ref{eq:2-1}). 

In order to apply the stability definition \ref{Def:Stab2} to the
class of policies specified above, we must define the time instant
at which a packet leaves the system. There are two ways to define
this instant. According to the first, a packet is considered to leave
the system when it is correctly decoded by the destination receiver.
While this definition make sense if one is interested in packet delivery
times, it does not capture the fact that a decoded packet may still
be needed for further encoding and decoding, in which case the packet
will keep occupying buffer space even after its correct decoding.
Also, the feedback information may need to be stored in the buffers
of the transmitter if needed for further encoding. To capture buffer
requirements we assume that each of the receivers has infinite buffers
where received packets are stored. We next introduce a second definition
of queue size, where we take into account the following.
\begin{enumerate}
\item Each transmission results in storing at most $n$ packets, one at
each receiver. These packets may be functions of ``native''
packets that have arrived exogenously at the transmitter, as well
as the feedback received at the transmitter. Hence, in this case packets may be generated
internally to the system during its operation. 
\item A packet stored at a receiver buffer departs when it is not needed
for further decoding.
\item A feedback packet is stored at the transmitter until it is not used
for further encoding.
\item A native packet departs from the receiver if a) it has been decoded
by the receiver to which it is destined and b) it is not used for
further encoding.
\end{enumerate}
If $Q_{D}^{\pi}\left(t\right)$ and $Q_{B}^{\pi}\left(t\right)$ respectively
are the sum of queue sizes under $\pi$ according to the previous
two definitions of packet departure time (Delay, Buffer), it holds,
$Q_{D}^{\pi}\left(t\right)\leq Q_{B}^{\pi}\left(t\right).$ Hence,
if $\mathcal{S}_{D}$ and $\mathcal{S}_{B}$ are respectively the
stability regions according to the two definitions, it holds 
\begin{equation}
\mathcal{S}_{B}\subseteq\mathcal{S}_{D}.\label{eq:cap24}
\end{equation}

\subsection*{Relation between Capacity and Stability Regions}

The distributed nature of the channel introduces some new issues that
must be addressed in order to apply the results of the previous sections.
Specifically, while the transmitter has full knowledge of the system
through the channel feedback, this is not the case for the receivers.
Transferring appropriate information to the receivers takes extra
slots which must be accounted for.

Note first that there are some differences in the information available
at the receivers in the definition of the two regions given above.
Specifically, in the capacity region definition, it is assumed that
the receivers know the number of packets at the transmitter when the
algorithm starts. On the other hand, when arrivals are stochastic,
this information cannot be assumed a priori and if needed it must be
communicated to the receivers. Also, in the capacity definition, all
receivers under any admissible coding know implicitly the instant
$t$ at which the decoding process stops. For the stochastic arrival
model, however, under a general evacuation policy, this may not be
the case. Note that the One-By-One policy $\pi_{O}$ does not need
the information regarding the number of packets at the transmitter
when the system starts. Also, an evacuation policy that is based on
$\pi_{O}$ can be easily modified to inform the receivers about the
end of the decoding process: when all packets to destination $i$
are transmitted, an empty slot is transmitted in the next slot allocated
to $i,$ informing all receivers of this event. Hence if the last
packet is delivered to the appropriate destination at time $t$, all
receivers will know at time $t+1$ that all packets are evacuated.
Note that (\ref{eq:2-1}) still holds under this modification. We
denote this modified policy as $\pi_{O}^{e}$.

Since it can be preagreed which evacuation policy to employ when a
given number of packets $\boldsymbol{k}$ is initially at the transmitter,
once that number is known
by all receivers, the employed evacuation policy is also known by
the receivers. 

In the following, we initially assume the following conditions (these
conditions will be removed later).
\begin{itemize}
\item When an evacuation policy starts, the number of packets at the transmitter
is known to the receivers.
\item An evacuation policy ensures that all receivers realize the end of
the evacuation process at some time $t$, which is defined as the
end of the evacuation process.
\end{itemize}
Under these conditions, the arguments of Lemma \ref{lem:0} apply
and hence $\bar{T}^{*}(\boldsymbol{k})$ is again subadditive (we
omit the subscript describing states since the system under discussion
has just a single state). Also, the arguments for (\ref{eq:5-1})
still hold (note that by placing the ``dummy'' packet in the argument
last in the transmitter queue corresponding to receiver $i,$ this
receiver knows that this packet contains no information and hence
decoding error does not occur). Hence Theorem \ref{lem:4-1} holds
for the current model.

We now claim that under the above stated assumptions,
\begin{lem}
\label{lem:16} It holds,
\[
\mathcal{C}=\mathcal{R}\triangleq\left\{ \boldsymbol{r}\geq\boldsymbol{0}:\hat{T}\left(\boldsymbol{r}\right)\leq1\right\} .
\]
\end{lem}
\begin{IEEEproof}
We first show that $\mathcal{R}\subseteq\mathcal{C}.$ For this, it
suffices to show that if for some $\boldsymbol{r}$ it holds $\hat{T}\left(\boldsymbol{r}\right)<1,$
then there is a sequence of codes $c_{l}=(2^{\left\lceil lr_{1}\right\rceil L},\ldots,2^{\left\lceil lr_{n}\right\rceil L},l)$
with $q_{l}^{e}\rightarrow_{l\rightarrow\infty}0.$ Select $\delta>0$
such that 
\begin{equation}
\hat{T}\left(\boldsymbol{r}\right)+3\delta<1.\label{eq:0i}
\end{equation}
In the following, we denote, for any positive integers $l$ and $l_0$,
$\alpha_l=\left\lfloor\frac{l}{l_0}\right\rfloor$ and $\beta_l=l\mod l_0$,
i.e.\
\[
l=\alpha_{l}l_{0}+\beta_{l},\:0\leq\beta_{l}<l_{0}
\]
It follows that
\begin{equation}
\left\lceil lr_{i}\right\rceil \leq\left(\alpha_{l}+1\right)\left\lceil l_{0}r_{i}\right\rceil .\label{eq:cap41}
\end{equation}

Select and fix $l_{0}$ large enough so that
\begin{equation}
\frac{\bar{T}^{*}\left(\left\lceil l_{0}\boldsymbol{r}\right\rceil \right)}{l_{0}}\leq\hat{T}\left(\boldsymbol{r}\right)+\delta.\label{eq:2-1i}
\end{equation}

Select an evacuation policy $\pi_{l_{0}}$ such that
\begin{equation}
\bar{T}_{\pi_{l_{0}}}\left(\left\lceil l_{0}\boldsymbol{r}\right\rceil \right)\leq\bar{T}^{*}\left(\left\lceil l_{0}\boldsymbol{r}\right\rceil \right)+l_{0}\delta,\label{eq:3i}
\end{equation}

Consider the following sequence of codes\textbf{ $\ensuremath{c_{l}}$
}for transmitting $\left\lceil l\boldsymbol{r}\right\rceil $ packets.

a) Use $\pi_{l_{0}}$ to transmit successively $\alpha_l+1$ batches
of $\left\lceil l_{0}\boldsymbol{r}\right\rceil $ packets (the last
batch may contain dummy packets) until they are decoded by all receivers.
Let $T_{\pi_{l_{0}}}^{j}\left(\left\lceil l_{0}\boldsymbol{r}\right\rceil \right)$
be the (random) time it takes to transmit the $j$-th batch, and 
\[
\tilde{T}_{\pi_{l_{0}}}^{l}\left(\left\lceil l_{0}\boldsymbol{r}\right\rceil \right)=\sum_{j=1}^{\alpha_l+1}T_{\pi_{l_{0}}}^{j}\left(\left\lceil l_{0}\boldsymbol{r}\right\rceil \right)
\]

b) If 
\[
\tilde{T}_{\pi_{l_{0}}}^{l}\left(\left\lceil l_{0}\boldsymbol{r}\right\rceil \right)\leq l
\]
 all packets are correctly decoded; else declare an error.

The probability of error for the sequence $c_{l}$ is computed as follows. Observing
that 
\[
\lim_{l\rightarrow\infty}\alpha_l=\infty,\:\lim_{l\rightarrow\infty}\frac{\beta_{l}}{\alpha_{l}}=0,
\]
and taking into account (\ref{eq:0i}), pick $\tilde{l}$ large enough
so that for all $l\geq\tilde{l}$ it holds, 
\begin{align}
\frac{\alpha_{l}}{\alpha_{l}+1}\left(1+\frac{\beta_{l}}{l_{0}\alpha_{l}}\right) & =\left(1-\frac{1}{\alpha_{l}+1}\right)\left(1+\frac{\beta_{l}}{l_{0}\alpha_{l}}\right)\nonumber \\
 & =1+\frac{\beta_{l}}{l_{0}\alpha_{l}}-\frac{1}{\alpha_{l}+1}\left(1+\frac{\beta_{l}}{l_{0}\alpha_{l}}\right)\nonumber \\
 & \geq1-\frac{1}{\alpha_{l}+1}\left(1+\frac{\beta_{l}}{l_{0}\alpha_{l}}\right)\nonumber \\
 & \geq\hat{T}\left(\boldsymbol{r}\right)+3\delta\label{eq:4i}
\end{align}
Then, 
\begin{align*}
q_{l}^{e} & =\Pr\left\{ \tilde{T}_{\pi_{l_{0}}}^{l}\left(\left\lceil l_{0}\boldsymbol{r}\right\rceil \right)>l\right\} \\
 & =\Pr\left\{ \sum_{j=1}^{\alpha_{l}+1}T_{\pi_{l_{0}}}^{j}\left(\left\lceil l_{0}\boldsymbol{r}\right\rceil \right)>\alpha_{l}l_{0}+\beta_{l}\right\} \\
 & =\Pr\left\{ \frac{\sum_{j=1}^{\alpha_{l}+1}\frac{T_{\pi_{l_{0}}}^{j}\left(\left\lceil l_{0}\boldsymbol{r}\right\rceil \right)}{l_{0}}}{\alpha_{l}+1}>\frac{\alpha_{l}}{\alpha_{l}+1}\left(1+\frac{\beta_{l}}{l_{0}\alpha_{l}}\right)\right\} \\
 & \leq\Pr\left\{ \frac{\sum_{j=1}^{\alpha_{l}+1}\frac{T_{\pi_{l_{0}}}^{j}\left(\left\lceil l_{0}\boldsymbol{r}\right\rceil \right)}{l_{0}}}{\alpha_{l}+1}>\hat{T}\left(\boldsymbol{r}\right)+3\delta\right\} {\rm \qquad by\quad(\ref{eq:4i})}\\
 & \leq\Pr\left\{ \left|\frac{\sum_{j=1}^{\alpha_{l}+1}\frac{T_{\pi_{l_{0}}}^{j}\left(\left\lceil l_{0}\boldsymbol{r}\right\rceil \right)}{l_{0}}}{\alpha_{l}+1}-\frac{\bar{T}_{\pi_{l_{0}}}\left(\left\lceil l_{0}\boldsymbol{r}\right\rceil \right)}{l_{0}}\right|>\hat{T}\left(\boldsymbol{r}\right)-\frac{\bar{T}_{\pi_{l_{0}}}\left(\left\lceil l_{0}\boldsymbol{r}\right\rceil \right)}{l_{0}}+3\delta\right\} \\
 & \leq\Pr\left\{ \left|\frac{\sum_{j=1}^{\alpha_{l}+1}\frac{T_{\pi_{l_{0}}}^{j}\left(\left\lceil l_{0}\boldsymbol{r}\right\rceil \right)}{l_{0}}}{\alpha_{l}+1}-\frac{\bar{T}_{\pi_{l_{0}}}\left(\left\lceil l_{0}\boldsymbol{r}\right\rceil \right)}{l_{0}}\right|>\hat{T}\left(\boldsymbol{r}\right)-\frac{\bar{T}^{*}\left(\left\lceil l_{0}\boldsymbol{r}\right\rceil \right)}{l_{0}}+2\delta\right\} {\rm \quad{\rm by}\:}(\ref{eq:3i})\\
 & \leq\Pr\left\{ \left|\frac{\sum_{j=1}^{\alpha_{l}+1}\frac{T_{\pi_{l_{0}}}^{j}\left(\left\lceil l_{0}\boldsymbol{r}\right\rceil \right)}{l_{0}}}{\alpha_{l}+1}-\frac{\bar{T}_{\pi_{l_{0}}}\left(\left\lceil l_{0}\boldsymbol{r}\right\rceil \right)}{l_{0}}\right|>\delta\right\} \quad{\rm by\;}(\ref{eq:2-1i}).
\end{align*}
Due to the memorylessness of the channel and the fact that the bits in
the packet contents are i.i.d, the random variables $T_{\pi_{l_{0}}}^{j}\left(\left\lceil l_{0}\boldsymbol{r}\right\rceil \right),\: j=1,2...$
are i.i.d. Using the fact that $\alpha_{l}\rightarrow_{l\rightarrow\infty}\infty$,
we conclude 
\[
\lim_{l\rightarrow\infty}\frac{\sum_{j=1}^{\alpha_{l}+1}\frac{T_{\pi_{l_0}}^{j}\left(\left\lceil l_{0}\boldsymbol{r}\right\rceil \right)}{l_{0}}}{\alpha_{l}+1}=\frac{\bar{T}_{\pi_{l_{0}}}\left(\left\lceil l_{0}\boldsymbol{r}\right\rceil \right)}{l_{0}}
\]
which implies that
\begin{align*}
\lim_{l\rightarrow\infty}q_{l}^{e} & =\lim_{l\rightarrow\infty}\Pr\left\{ \left|\frac{\sum_{j=1}^{\alpha_{l}+1}\frac{T_{\pi_{l_{0}}}^{j}\left(\left\lceil l_{0}\boldsymbol{r}\right\rceil \right)}{l_{0}}}{\alpha_{l}+1}-\frac{\bar{T}_{\pi_{l_{0}}}\left(\left\lceil l_{0}\boldsymbol{r}\right\rceil \right)}{l_{0}}\right|>\delta\right\} =0.
\end{align*}

Next we show that $\mathcal{C}\subseteq\mathcal{R}.$ Assume that
$\boldsymbol{r}\in\mathcal{C}$ so that there is a sequence of coding
algorithms $c_{l}$ with rate $\boldsymbol{r}$ whose error probability
approaches zero in the limit as $l\to\infty$. We then construct an evacuation
policy $\pi_{l}$ for evacuating $\left\lceil l\boldsymbol{r}\right\rceil$
packets as follows.

a) For $\epsilon>0,$ select $l$ so that $q_{l}^{e}<\epsilon$.

b) Follow the steps of $c_{l}$ for the first $l$ slots. 

c) If all receivers decoded correctly, leave slot $l+1$ empty, thus
signaling to all receivers the end of the decoding process.

e) Else (i.e., if any of the receivers makes an error), send a dummy
packet in slot $l+1$ (thus informing the receivers that decoding
continues) and resend all the $\left\lceil l\boldsymbol{r}\right\rceil $
packets using the one-by-one policy $\pi_{O}.$

Note that, since the transmitter knows $g_i$ and, through the received feedback, the sequence received by $i$, it knows whether a
receiver makes an error and hence the third step above is implementable.

We compute the average evacuation time of $\pi_{l}$ as follows. Let
$\mathcal{E}$ be the event that all destinations have decoded the
packets in $l$ slots. Then, since on $\mathcal{E}^{c}$ it holds
\[
T_{\pi_{l}}\left(\left\lceil l\boldsymbol{r}\right\rceil \right)=l+T_{\pi_{O}}\left(\left\lceil l\boldsymbol{r}\right\rceil \right)+1,
\]
 $T_{\pi_{O}}\left(\left\lceil l\boldsymbol{r}\right\rceil \right)$
is independent of $\mathcal{E}^{c}$, and by choice $\Pr\left\{ \mathcal{E}^{c}\right\} =q_{l}^{e}<\epsilon,$
therefore we have
\begin{align*}
\mathbb{E}\left[T_{\pi_{l}}\left(\left\lceil l\boldsymbol{r}\right\rceil \right)1_{\mathcal{E}^{c}}\right] & =l\Pr\left\{ \mathcal{E}^{c}\right\} +\Pr\left\{ \mathcal{E}^{c}\right\} (\bar{T}_{\pi_{O}}+1)\\
 & \leq l\epsilon+\epsilon\left(C_{1}\sum_{i=1}^{n}\left\lceil lr_{i}\right\rceil +C_{0}+1\right).
\end{align*}
Taking into account that $T_{\pi_{l}}\left(\left\lceil l\boldsymbol{r}\right\rceil \right)=l+1$
on $\mathcal{E},$ 
\begin{align*}
\mathbb{E}\left[T_{\pi_{l}}\left(\left\lceil l\boldsymbol{r}\right\rceil \right)\right] & =\mathbb{E}\left[T_{\pi_{l}}\left(\left\lceil l\boldsymbol{r}\right\rceil \right)1_{\mathcal{E}}\right]+\mathbb{E}\left[T_{\pi_{l}}\left(\left\lceil l\boldsymbol{r}\right\rceil \right)1_{\mathcal{E}^{c}}\right]\\
 & \leq l+1+\epsilon\left(l+C_{1}\sum_{i=1}^{n}\left\lceil lr_{i}\right\rceil +C_{0}+1\right)
\end{align*}
Hence, 
\begin{align*}
\frac{\bar{T}_{A}^{*}\left(\left\lceil l\boldsymbol{r}\right\rceil \right)}{l} & \leq\frac{\bar{T}_{\pi_{l}}\left(\left\lceil l\boldsymbol{r}\right\rceil \right)}{l}\\
 & \leq1+\frac{1}{l}+\epsilon\cdot\frac{l+C_{1}\sum_{i=1}^{n}\left\lceil lr_{i}\right\rceil +C_{0}+1}{l}
\end{align*}
Considering the limit as $l\to\infty$, we obtain, 
\[
\hat{T}_{A}\left(\boldsymbol{r}\right)\leq1+\epsilon\left(C_{1}\sum_{i=1}^{n}r_{i}+1\right)
\]
and since $\epsilon$ is arbitrary we conclude 
\[
\hat{T}_{A}\left(\boldsymbol{r}\right)\leq1.
\]

\end{IEEEproof}
It remains to relate $\mathcal{R}$ to $\mathcal{S}_{D}$ and $\mathcal{S}_{B}$
under the current model. Revisit the proof of Theorem \ref{lem:3},
and use a policy $\pi_{0}\in\mathcal{S}_{D}$ for the first $l$ slots.
It all packets are decoded correctly by slot $l$, leave slot $l+1$ empty,
thus informing all receivers of successful decoding. Else send a dummy
packet in slot $l+1$ and afterwards apply the One-By-One policy $\pi_{O}$
as policy $\pi_{h}$ in the proof to evacuate the remaining packets.
With these modifications, the proof can be used to show that 
\begin{equation}
\mathcal{S}_{D}\subseteq\mathcal{R}.\label{eq:cap30}
\end{equation}

We now consider the implementation of the Epoch Based policy $\pi_{\epsilon}$
under the current model. This policy selects a particular evacuation
policy for each epoch, which is a function of the number of packets
$\boldsymbol{k}$ at the beginning of the epoch. In order to implement
$\pi_{\epsilon}$ in the current model, the receivers must generally
know $\boldsymbol{k}$ at the beginning of an epoch. The transfer
of information about the number $\boldsymbol{k}$ is done by transmitting
$O\left(\sum_{i=1}^{n}\log\left(k_{i}+1\right)\right)$ packets (for example,
using the One-by-One policy $\pi_{O}$) and hence the average number of slots to
achieve this transfer is
$O\left(\sum_{i=1}^{n}\log\left(k_{i}+1\right)\right)$.
This increases the length of the evacuation period but since the increase
is logarithmic in the number of packets, it does not affect the stability
arguments. Note also that once an epoch ends, all $\boldsymbol{k}$
packets, as well as the feedback information and the packets stored
at the receivers can be discarded since they are not used for further
decoding by $\pi_{\epsilon}.$ Hence we conclude that 
\begin{equation}
\mathcal{R}\subseteq\mathcal{S}_{B}\label{eq:cap31}
\end{equation}
Taking into account (\ref{eq:cap24}), (\ref{eq:cap30}), (\ref{eq:cap31})
we finally conclude,
\begin{thm}
\label{thm:cap17}It holds,
\[
\mathcal{C}=\mathcal{R}=\mathcal{S}_{B}=\mathcal{S}_{D}.
\]

\end{thm}


\appendices{}

\section{Proof of Theorem \ref{lem:4-1}}

An analogous to Theorem \ref{lem:4-1} has been derived in~\cite{BiOs08}
for subadditive functions defined on $\mathbb{R}^{n}.$ The extension
of Critical Evacuation Time Function to $\mathbb{R}_{0}^{n}$ given
in (\ref{eq:8-f}) is not necessarily subadditive and hence we need
different arguments to show the result, albeit using similar ideas. 

Let $f\left(\boldsymbol{k}\right):\mathbb{N}_{0}^{n}\rightarrow\mathbb{R}_{0}$
be a subadditive function. Let $\mathcal{U}$ be the set of $n$-dimensional
vectors whose coordinates are either zero or one, and define, 
\[
U=\max_{\boldsymbol{u}\in\mathcal{U}}f\left(\boldsymbol{u}\right).
\]
 We will need the following lemma. 
\begin{lem}
\label{lem:9}For any $\boldsymbol{k}\in\mathbb{N}_{0}^{n}-\left\{ \boldsymbol{0}\right\} ,$
it holds 
\[
f\left(\boldsymbol{k}\right)\leq U\max_{i}k_{i}.
\]
\end{lem}
\begin{IEEEproof}
Assume without loss of generality that for some $c\leq n,$ $0<k_{1}\leq k_{2}\leq...\leq k_{c}$
and, in case $c<n,$ then $k_{c+1}=\ldots=k_n=0$. Write, 
\[
\boldsymbol{k}=\left[\begin{array}{c}
k_{1}\\
\vdots\\
k_{n}
\end{array}\right]=\sum_{i=1}^{c}\left(k_{i}-k_{i-1}\right)\boldsymbol{u}_{i},
\]
where $k_{0}=0$ and 
\[
u_{i,j}=\begin{cases}
0 & {\rm if}\:\: i>1\quad{\rm and}\quad j=1,...,i-1,\\
1 & {\rm if\:\:}j=i,...,c\\
0 & {\rm if\:\:}j>c
\end{cases}
\]
By subadditivity we have, 
\begin{align*}
f\left(\boldsymbol{k}\right) & \leq\sum_{i=1}^{c}\left(k_{i}-k_{i-1}\right)f\left(\boldsymbol{u}_{i}\right)\\
 & \leq Uk_{c}
\end{align*}

\end{IEEEproof}
Next we extend the definition of $f\left(\boldsymbol{k}\right)$ to
$\mathbb{R}_{0}^{n}$ by defining 
\[
f\left(\boldsymbol{r}\right)=f\left(\left\lceil \boldsymbol{r}\right\rceil \right),\:\boldsymbol{r}\in\mathbb{R}_{0}^{n}.
\]

We then have the following theorem.
\begin{thm}
\label{thm:Th8}For any $\boldsymbol{r}\in\mathbb{R}_{0}^{n}$, the
limit function 
\begin{equation}
\hat{f}\left(\boldsymbol{r}\right)=\lim_{t\rightarrow\infty}\frac{f\left(t\boldsymbol{r}\right)}{t}\label{eq:18}
\end{equation}
 exists, is finite and positively homogeneous. \end{thm}
\begin{IEEEproof}
Assume without loss of generality that $r_{1}\geq r_{2}\geq...\geq r_{n}.$
If $r_{1}=0$ then $\boldsymbol{r}=\boldsymbol{0}$ and (\ref{eq:18})
is obvious. Assume next that for some $c$, $1\leq c\leq n,$ $r_{c}>0$
and $r_{c+1}=0$. For consistency define $r_{n+1}=0$. 

Let $\epsilon>0$ and $\beta=\liminf_{t\rightarrow\infty}f\left(t\boldsymbol{r}\right)/t\geq0$.
Using Lemma \ref{lem:9} we have, 
\begin{align*}
\frac{f\left(t\boldsymbol{r}\right)}{t} & =\frac{f\left(\left\lceil t\boldsymbol{r}\right\rceil \right)}{t}\\
 & \leq U\frac{\max_{i}\left\{ \left\lceil tr_{i}\right\rceil \right\} }{t}\\
 & <U\frac{\max_{i}\left\{ tr_{i}\right\} +1}{t}\\
 & =U\left(\max_{i}\left\{ r_{i}\right\} +\frac{1}{t}\right)
\end{align*}
Hence, $\beta<\infty.$

To show existence of the limit in (\ref{eq:18}), it suffices to show
that 
\begin{equation}
\limsup_{t\rightarrow\infty}\frac{f\left(t\boldsymbol{r}\right)}{t}\leq\beta+\delta(\epsilon),\label{eq:19-L}
\end{equation}
where $\lim_{\epsilon\rightarrow0}\delta\left(\epsilon\right)=0.$ 

By definition of $\beta,$ there are infinitely many $t$, such that
$f\left(t\boldsymbol{r}\right)/t\leq\beta+\epsilon$. Since we also
have 
\begin{equation}
r_{i}\leq\frac{\left\lceil tr_{i}\right\rceil }{t}<r_{i}+\frac{1}{t},\label{eq:20-h1}
\end{equation}
we can pick $t_{0}$ large enough so that the following inequalities
hold. 
\begin{equation}
\frac{f\left(t_{0}\boldsymbol{r}\right)}{t_{0}}\leq\beta+\epsilon,\label{eq:21-h2}
\end{equation}
\begin{equation}
r_{i}\leq\frac{\left\lceil t_{0}r_{i}\right\rceil }{t_{0}}<r_{i}+\epsilon,\:\: i=1,...,k.\label{eq:20-h}
\end{equation}

Using Euclidean division, write for $i=1,...,c$ 
\begin{equation}
\left\lceil tr_{i}\right\rceil =l_{t,i}\left\lceil t_{0}r_{i}\right\rceil +\upsilon_{t,i},\:0\leq\upsilon_{t,i}\leq\left\lceil t_{0}r_{i}\right\rceil -1\label{eq:20-L-1}
\end{equation}
 If $c<n,$ define also, 
\begin{equation}
l_{t,i}=\upsilon_{t,i}=0,\: i=c+1,...,n\label{eq:24-i}
\end{equation}
We then have, 
\begin{align}
f\left(t\boldsymbol{r}\right) & =f\left(\left\lceil t\boldsymbol{r}\right\rceil \right)\nonumber \\
 & =f\left(l_{t,1}\left\lceil t_{0}r_{1}\right\rceil +\upsilon_{t,1},...,l_{t,n}\left\lceil t_{0}r_{n}\right\rceil +\upsilon_{t,n}\right)\nonumber \\
 & \leq f\left(l_{t,1}\left\lceil t_{0}r_{1}\right\rceil ,...,l_{t,n}\left\lceil t_{0}r_{n}\right\rceil \right)+f\left(\boldsymbol{\upsilon}_{t}\right)\:\:{\rm by\: subadditivity}\label{eq:20-L}
\end{align}
Next, write 
\[
\left[\begin{array}{c}
l_{t,1}\left\lceil t_{0}r_{1}\right\rceil \\
\vdots\\
l_{t,n}\left\lceil t_{0}r_{n}\right\rceil 
\end{array}\right]=\sum_{j=1}^{c}\left(l_{t,j}-l_{t,j-1}\right)\boldsymbol{v}_{j},
\]
where $l_{t,0}=0$ and the $i$th coordinate of $\boldsymbol{v}_{j}$,
$v_{j,i},$ is defined for $1\leq j\leq c$ as, 
\begin{equation}
v_{j,i}=\begin{cases}
0 & {\rm if}\:\: j\neq1\quad{\rm and}\quad i=1,...j-1,\\
\left\lceil t_{0}r_{i}\right\rceil  & {\rm if\:\:}i=j,...,n
\end{cases}\label{eq:26-i}
\end{equation}
Notice that since $r_{j}\geq r_{j+1},$ it holds, $l_{t,j-1}\leq l_{t,j},\:1\leq j\leq c.$
Using subadditivity, we then have from (\ref{eq:20-L}),
\[
f\left(t\boldsymbol{r}\right)\leq\sum_{j=1}^{c}\left(l_{t,j}-l_{t,j-1}\right)f\left(\boldsymbol{v}_{j}\right)+f\left(\boldsymbol{\upsilon}_{t}\right)
\]
Hence, 
\begin{align}
\frac{f\left(t\boldsymbol{r}\right)}{t} & \leq\sum_{j=1}^{c}\frac{\left(l_{t,j}-l_{t,j-1}\right)t_{0}}{t}\frac{f\left(\boldsymbol{v}_{j}\right)}{t_{0}}+\frac{f\left(\boldsymbol{\upsilon}_{t}\right)}{t}\nonumber \\
 & =\frac{l_{t,1}t_{0}}{t}\frac{f\left(t_{0}\boldsymbol{r}\right)}{t_{0}}+\sum_{j=2}^{c}\frac{\left(l_{t,j}-l_{t,j-1}\right)t_{0}}{t}\frac{f\left(\boldsymbol{v}_{j}\right)}{t_{0}}+\frac{f\left(\boldsymbol{\upsilon}_{t}\right)}{t}\label{eq:25-3}
\end{align}
By (\ref{eq:20-L-1}), (\ref{eq:24-i}), $\boldsymbol{\upsilon}_{t}$
takes a finite number of values, hence $f\left(\boldsymbol{\upsilon}_{t}\right)$
is a bounded sequence, and
\[
\lim_{t\rightarrow\infty}\frac{f\left(\boldsymbol{\upsilon}_{t}\right)}{t}=0.
\]
Also, from (\ref{eq:20-h1}), (\ref{eq:20-h}) and (\ref{eq:20-L-1})
we have for $1\leq i\leq c,$ 
\[
r_{i}\leq\frac{\left\lceil tr_{i}\right\rceil }{t}=\frac{l_{t,i}t_{0}}{t}\frac{\left\lceil t_{0}r_{i}\right\rceil }{t_{0}}+\frac{\upsilon_{t,i}}{t}<\frac{l_{t,i}t_{0}}{t}\left(r_{i}+\epsilon\right)+\frac{\upsilon_{t,i}}{t},
\]
\[
r_{i}+\frac{1}{t}>\frac{\left\lceil tr_{i}\right\rceil }{t}=\frac{l_{ti}t_{0}}{t}\frac{\left\lceil t_{0}r_{i}\right\rceil }{t_{0}}+\frac{\upsilon_{ti}}{t}\geq\frac{l_{ti}t_{0}}{t}r_{i}+\frac{\upsilon_{t,i}}{t},
\]
 hence, using the fact that $\boldsymbol{\upsilon}_{t}$ is a bounded
sequence, we conclude 
\[
1-\frac{\epsilon}{r_{c}}\leq1-\frac{\epsilon}{r_{i}}\leq\frac{r_{i}}{r_{i}+\epsilon}\leq\liminf_{t\rightarrow\infty}\frac{l_{t,i}t_{0}}{t}\leq\limsup\frac{l_{t,i}t_{0}}{t}\leq1.
\]
Taking into account the latter inequalities and (\ref{eq:21-h2})
we have from (\ref{eq:25-3}),
\begin{align*}
\limsup_{t\rightarrow\infty}\frac{f\left(t\boldsymbol{r}\right)}{t} & \leq\left(\beta+\epsilon\right)\limsup_{t\rightarrow\infty}\frac{l_{t,1}t_{0}}{t}+\sum_{j=2}^{c}\left(\limsup_{t\rightarrow\infty}\frac{l_{t,j}t_{0}}{t}-\liminf_{t\rightarrow\infty}\frac{l_{t,j-1}t_{0}}{t}\right)\frac{f\left(\boldsymbol{v}_{j}\right)}{t_{0}}\\
 & \leq\beta+\epsilon+\frac{\epsilon}{r_{c}}\sum_{j=2}^{c}\frac{f\left(\boldsymbol{v}_{j}\right)}{t_{0}}\\
 & \leq\beta+\epsilon+\frac{\epsilon}{r_{c}}Uc\frac{\max_{i}\left\lceil t_{0}r_{i}\right\rceil }{t_{0}}\:\:{\rm by\: Lemma\:}\ref{lem:9}\:{\rm and\;(}\ref{eq:26-i})\\
 & \leq\beta+\epsilon+\frac{\epsilon}{r_{c}}Un\left(r_{1}+\epsilon\right)\:\:{\rm by\:(\ref{eq:20-h}})
\end{align*}
Hence (\ref{eq:19-L}) holds with $\delta(\epsilon)=\epsilon+\frac{\epsilon}{r_{c}}Un\left(r_{1}+\epsilon\right)$.

Positive homogeneity follows immediately since for $\alpha\geq0$,
\[
\hat{f}\left(\alpha\boldsymbol{r}\right)=\lim_{t\rightarrow\infty}\frac{f\left(t\alpha\boldsymbol{r}\right)}{t}=\alpha\lim_{t\rightarrow\infty}\frac{f\left(t\alpha\boldsymbol{r}\right)}{\alpha t}=\alpha\hat{f}\left(\boldsymbol{r}\right).
\]

\end{IEEEproof}
The next lemma is needed to establish further properties of $f\left(\boldsymbol{k}\right)$
in Theorem \ref{thm:6} below.
\begin{lem}
\label{lem:5-1}Let a subadditive function $f\left(\boldsymbol{k}\right),\:\boldsymbol{k}\in\mathbb{N}_{0}^{n}$
satisfy 
\begin{equation}
f\left(\boldsymbol{k}\right)-f\left(\boldsymbol{k}+\boldsymbol{e}_{i}\right)\leq D_{0}\label{eq:28}
\end{equation}
Then the following holds with $D=\max\left\{ f\left(\boldsymbol{e}_{1}\right),...,f\left(\boldsymbol{e}_{n}\right),D_{0}\right\} .$
\begin{equation}
\left|f\left(\boldsymbol{k}\right)-f\left(\boldsymbol{k}+\boldsymbol{e}_{i}\right)\right|\leq D,\:{\rm for\: all\;}i=1,..,n.\label{eq:11-1}
\end{equation}
\begin{align}
\left|f\left(\boldsymbol{k}\right)-f\left(\boldsymbol{m}\right)\right| & \leq D\sum_{i=1}^{n}\left|k_{i}-m_{i}\right|\label{eq:12-1}\\
\left|f\left(\boldsymbol{r}\right)-f\left(\boldsymbol{s}\right)\right| & <D\sum_{i=1}^{n}\left|r_{i}-s_{i}\right|+nD\label{eq:13-1}
\end{align}
\end{lem}
\begin{IEEEproof}
By subadditivity, 
\[
f\left(\boldsymbol{k}+\boldsymbol{e}_{i}\right)\leq f\left(\boldsymbol{k}\right)+f\left(\boldsymbol{e}_{i}\right)
\]
hence, 
\[
f\left(\boldsymbol{k}+\boldsymbol{e}_{i}\right)-f\left(\boldsymbol{k}\right)\leq\max_{i}\bar{T}^{*}\left(\boldsymbol{e}_{i}\right)\doteq D_{1}
\]
Taking into account (\ref{eq:28}) we conclude, 
\[
\left|f\left(\boldsymbol{k}+\boldsymbol{e}_{i}\right)-f\left(\boldsymbol{k}\right)\right|\leq\max\left\{ D_{1},D_{0}\right\} \doteq D
\]
 which shows (\ref{eq:11-1}).

To show (\ref{eq:12-1}) we use backward induction on the number $c$
of coordinates of $\boldsymbol{k},\:\boldsymbol{m}$ that are equal.
If $c=n$ then clearly \eqref{eq:12-1} holds. Let \eqref{eq:12-1}
hold for $c\leq n$ and assume without loss of generality that $k_{i}=m_{i},\: i=1,...,c-1$
and $k_{i}\neq m_{i},\: i\geq c,\: k_{c}>m_{c}$. We then have 
\begin{align*}
\left|f\left(\boldsymbol{k}\right)-f\left(\boldsymbol{m}\right)\right| & =\left|f\left(\boldsymbol{k}\right)-f\left(k_{1},...k_{c-1},m_{c},k_{c+1},...,k_{n}\right)+f\left(k_{1},...k_{c-1},m_{c},k_{c+1},...,k_{n}\right)-f\left(\boldsymbol{m}\right)\right|\\
 & \leq\left|f\left(\boldsymbol{k}\right)-f\left(k_{1},...k_{c-1},m_{c},k_{c+1},...,k_{n}\right)\right|+\left|f\left(m_{1},...m_{c-1},m_{c},k_{c+1},...,k_{n}\right)-f\left(\boldsymbol{m}\right)\right|\\
 & \leq\left|f\left(\boldsymbol{k}\right)-f\left(k_{1},...k_{c-1},m_{c},k_{c+1},...,k_{n}\right)\right|+D\sum_{i=c+1}^{n}\left|k_{i}-m_{i}\right|\:\:{\rm by\: the\: inductive\: hypothesis}.
\end{align*}
Now, write 
\begin{align*}
\left|f\left(\boldsymbol{k}\right)-f\left(k_{1},...k_{c-1},m_{c},k_{c+1},...,k_{n}\right)\right| & =\left|\sum_{i=0}^{k_{c}-m_{c}-1}f\left(k_{1},...k_{c-1},m_{c}+i+1,k_{c+1},...,k_{n}\right)-f\left(k_{1},...k_{c-1},m_{c}+i,k_{c+1},...,k_{n}\right)\right|\\
 & \leq\sum_{i=0}^{k_{c}-m_{c}-1}\left|f\left(k_{1},...k_{c-1},m_{c}+i+1,k_{c+1},...,k_{n}\right)-f\left(k_{1},...k_{c-1},m_{c}+i,k_{c+1},...,k_{n}\right)\right|\\
 & \leq\sum_{i=0}^{k_{c}-m_{c}-1}D\:\:\:{\rm by\:(}\ref{eq:11-1})\\
 & =D\left|k_{c}-m_{c}\right|
\end{align*}
and hence, 
\[
\left|f\left(\boldsymbol{k}\right)-f\left(\boldsymbol{m}\right)\right|\leq D\sum_{i=c}^{n}\left|k_{i}-m_{i}\right|=D\sum_{i=1}^{n}\left|k_{i}-m_{i}\right|\:\:{\rm since\;}k_{i}=m_{i},\: i=1,...,c-1
\]
i.e., the inductive hypothesis holds for $c-1$ as well. 

Finally, for (\ref{eq:13-1}), write 
\begin{align*}
\left|f\left(\boldsymbol{r}\right)-f\left(\boldsymbol{s}\right)\right| & =\left|f\left(\left\lceil \boldsymbol{r}\right\rceil \right)-f\left(\left\lceil \boldsymbol{s}\right\rceil \right)\right|\\
 & \leq D\sum_{i=1}^{n}\left|\left\lceil r_{i}\right\rceil -\left\lceil s_{i}\right\rceil \right|\:\:{\rm by\;}(\ref{eq:12-1})\\
 & <D\sum_{i=1}^{n}\left|r_{i}-s_{i}\right|+Dn,\:{\rm since\;}\left|\left\lceil r_{i}\right\rceil -\left\lceil s_{i}\right\rceil \right|<\left|r_{i}-s_{i}\right|+1
\end{align*}

\end{IEEEproof}
The next theorem provides further useful properties of $\hat{f}\left(\boldsymbol{r}\right)$
under condition (\ref{eq:28}).
\begin{thm}
\label{thm:6}If a subadditive function $f\left(\boldsymbol{k}\right),\:\boldsymbol{k}\in\mathbb{N}_{0}^{n}$
satisfies (\ref{eq:28}), then the limit function 
\[
\hat{f}\left(\boldsymbol{r}\right)=\lim_{t\rightarrow\infty}\frac{f\left(t\boldsymbol{r}\right)}{t}
\]
 is subadditive, convex, Lipschitz continuous, i.e., it holds 
\[
\left|\hat{f}\left(\boldsymbol{r}\right)-\hat{f}\left(\boldsymbol{s}\right)\right|\leq D\sum_{i=1}^{n}\left|r_{i}-s_{i}\right|.
\]
 and for any sequence $\boldsymbol{r}_{t}\in\mathbb{R}_{0}^{n}$ such
that
\[
\lim_{t\rightarrow\infty}\boldsymbol{r}_{t}=\boldsymbol{\lambda}<\boldsymbol{\infty,}
\]
it holds 
\begin{equation}
\lim_{t\rightarrow\infty}\frac{f\left(t\boldsymbol{r}_{t}\right)}{t}=\hat{f}\left(\boldsymbol{\lambda}\right).\label{eq:32}
\end{equation}
 \end{thm}
\begin{IEEEproof}
To show subadditivity, we proceed as follows. Since for any $a,\: b$
it holds 
\[
\left\lceil a+b\right\rceil +x=\left\lceil a\right\rceil +\left\lceil b\right\rceil \:{\rm for\: some\:}x=0,1,2,
\]
 we write
\[
\left\lceil t\left(\boldsymbol{r}_{1}+\boldsymbol{r}_{2}\right)\right\rceil +\boldsymbol{x}=\left\lceil t\boldsymbol{r}_{1}\right\rceil +\left\lceil t\boldsymbol{r}_{2}\right\rceil .
\]
Also, by (\ref{eq:12-1}) 
\begin{align*}
f\left(\left\lceil t\left(\boldsymbol{r}_{1}+\boldsymbol{r}_{2}\right)\right\rceil \right)-f\left(\left\lceil t\left(\boldsymbol{r}_{1}+\boldsymbol{r}_{2}\right)\right\rceil +\boldsymbol{x}\right) & \leq D\sum_{i=}^{n}x_{i}\\
 & \leq2nD
\end{align*}
 Hence, 
\begin{align*}
f\left(t\left(\boldsymbol{r}_{1}+\boldsymbol{r}_{2}\right)\right)-2nD & \leq f\left(\left\lceil t\left(\boldsymbol{r}_{1}+\boldsymbol{r}_{2}\right)\right\rceil +\boldsymbol{x}\right)\\
 & =f\left(\left\lceil t\boldsymbol{r}_{1}\right\rceil +\left\lceil t\boldsymbol{r}_{2}\right\rceil \right)\\
 & \leq f\left(t\boldsymbol{r}_{1}\right)+f\left(t\boldsymbol{r}_{2}\right)
\end{align*}
Dividing the last inequality by $t$ and taking limits shows that
$\hat{f}\left(\boldsymbol{r}_{1}+\boldsymbol{r}_{2}\right)\leq\hat{f}\left(\boldsymbol{r}_{1}\right)+\hat{f}\left(\boldsymbol{r}_{2}\right)$. 

Convexity follows easily from positive homogeneity and subadditivity,
\begin{align*}
\hat{f}\left(p\boldsymbol{r}_{1}+(1-p)\boldsymbol{r}_{2}\right) & \leq\hat{f}\left(p\boldsymbol{r}_{1}\right)+\hat{f}\left((1-p)\boldsymbol{r}_{2}\right)\\
 & =p\hat{f}\left(\boldsymbol{r}_{1}\right)+(1-p)\hat{f}\left(\boldsymbol{r}_{2}\right).
\end{align*}

Lipschitz continuity follows easily as well from (\ref{eq:13-1}) by
replacing $\boldsymbol{r},\:\boldsymbol{s}$ with $t\boldsymbol{r},\: t\boldsymbol{s},$
dividing by $t$ and taking limits.

Finally let 
\[
\lim_{t\rightarrow\infty}\boldsymbol{r}_{t}=\boldsymbol{\lambda}<\boldsymbol{\infty}
\]
Using (\ref{eq:13-1}) write
\begin{align*}
\left|\frac{f\left(t\boldsymbol{r}_{t}\right)}{t}-\hat{f}\left(\boldsymbol{\lambda}\right)\right| & =\left|\frac{f\left(t\boldsymbol{r}_{t}\right)}{t}-\frac{f\left(t\boldsymbol{\lambda}\right)}{t}+\frac{f\left(t\boldsymbol{\lambda}\right)}{t}-\hat{f}\left(\boldsymbol{\lambda}\right)\right|\\
 & \leq\left|\frac{f\left(t\boldsymbol{r}_{t}\right)-f\left(t\boldsymbol{\lambda}\right)}{t}\right|+\left|\frac{f\left(t\boldsymbol{\lambda}\right)}{t}-\hat{f}\left(\boldsymbol{\lambda}\right)\right|\\
 & \leq D\sum_{i=1}^{n}\left|r_{t,i}-\lambda_{i}\right|+\frac{nD}{t}+\left|\frac{f\left(t\boldsymbol{\lambda}\right)}{t}-\hat{f}\left(\boldsymbol{\lambda}\right)\right|
\end{align*}
Taking limits in the last inequality shows (\ref{eq:32}).
\end{IEEEproof}
Theorem \ref{lem:4-1} will follow directly from Theorems \ref{thm:Th8},
\ref{thm:6} if we verify that the critical evacuation time function
satisfies (\ref{eq:28}). But this follows easily from (\ref{eq:5-3})
since
\begin{align*}
\bar{T}^{*}\left(\boldsymbol{k}\right)-\bar{T}^{*}\left(\boldsymbol{k}+\boldsymbol{e}_{i}\right) & =\max_{s}\bar{T}_{s}^{*}\left(\boldsymbol{k}\right)-\max_{i}\bar{T}_{s}^{*}\left(\boldsymbol{k}+\boldsymbol{e}_{i}\right)\\
 & \leq\max_{s}\left\{ \bar{T}_{s}^{*}\left(\boldsymbol{k}\right)-\bar{T}_{s}^{*}\left(\boldsymbol{k}+\boldsymbol{e}_{i}\right)\right\} \\
 & \leq D_{0,}{\rm \: by\:(\ref{eq:5-3})}.
\end{align*}

\section{Proof of Theorem \ref{thm:15}}

In the discussion that follows we use the terminology and related
results in \cite{chung}. For $x,y\in\mathcal{G}$,
if $x$ \emph{leads} to $y,$ we write $x\rightsquigarrow y$ and
if $x$ \emph{communicates} with $y,$ $x\leftrightsquigarrow y.$
A Markov Chain with countable state space $\mathcal{G}$ is called
irreducible if all states in $\mathcal{G}$ belong to the same \emph{essential
}class, i.e., all states communicate with each other.

The proof of stability of the Epoch Based Policy is based on the following
theorem, see \cite{popov, issacson}.
\begin{thm}
\label{thm:Meyn}Let $\left\{ X_{m}\right\} _{m=1}^{\infty}$ be a
homogeneous, irreducible and aperiodic Markov Chain with countable
state space $\mathcal{G}$. Let $v(x)$ be a nonnegative real function
defined on the state space (Lyapunov function). If there exists a
finite set $\mathcal{A}\subseteq\mathcal{G}$ such that $v(x)\geq\epsilon>0,$
$x\in\mathcal{A}^{c}=\mathcal{G}-\mathcal{A},$
\begin{equation}
\mathbb{E}\left[v(X_{2})\left|X_{1}=x\right.\right]<\infty,\: x\in\mathcal{A},\label{eq:19-l}
\end{equation}
 and for some $\delta,\:1\geq\delta>0,$
\begin{equation}
\mathbb{E}\left[v(X_{2})\left|X_{1}=x\right.\right]\leq\left(1-\delta\right)v(x),\: x\in\mathcal{A}^{c},\label{eq:20-l}
\end{equation}
 then the Markov Chain is geometrically ergodic (positive recurrent)
and $\mathbb{E}\left[v\left(\overset{\smallfrown}{X}\right)\right]<\infty$,
where $\overset{\smallfrown}{X}$ has the steady-state distribution
of $\left\{ X_{m}\right\} _{m=1}^{\infty}$. 
\end{thm}
For the general model under consideration in the current work, irreducibility
and aperiodicity may not hold. Hence, we need some preparatory work
to use Theorem \ref{thm:Meyn}. The following lemma will be useful.
\begin{lem}
\label{lem:13}Let $\left\{ X_{m}\right\} _{m=1}^{\infty}$ be a homogeneous
Markov Chain, not necessarily irreducible and/or aperiodic. 

a) With the notation of Theorem \ref{thm:Meyn}, conditions (\ref{eq:19-l})
and (\ref{eq:20-l}) imply 
\begin{equation}
\mathbb{E}\left[v(X_{2})\left|X_{1}=x\right.\right]\leq U+\left(1-\delta\right)v(x)\:{\rm for\: all\;}x\in\mathcal{G},\label{eq:22-l}
\end{equation}
where $U=\max_{x\in\mathcal{A}}\mathbb{E}\left[v(X_{2})\left|X_{1}=x\right.\right].$ 

b) Conversely, if $v\left(x\right)\geq0$ and there are constants
$U>0$, $\delta,\:\delta_{1},$ $0<\delta_{1}<\delta\leq1,$ and a
finite set $\mathcal{B}$ such that (\ref{eq:22-l}) holds and
\begin{equation}
\frac{U}{v\left(x\right)}\leq\delta_{1}\:{\rm for\: all\:}x\in\mathcal{B}^{c},\label{eq:21-l}
\end{equation}
 then (\ref{eq:19-l}) and (\ref{eq:20-l}) hold with $\mathcal{A}\leftarrow\mathcal{B}$
and $\delta\leftarrow\delta-\delta_{1}.$ 

c) If (\ref{eq:22-l}) holds, then for $m\ge 2$,
\begin{equation}
\mathbb{E}\left[v(X_{m})\left|X_{1}=x\right.\right]\leq\frac{U}{\delta}+\left(1-\delta\right)^{m}v\left(x\right).\label{eq:23-l}
\end{equation}
\end{lem}
\begin{IEEEproof}
It is clear that (\ref{eq:19-l}) and (\ref{eq:20-l}) imply (\ref{eq:22-l}).
Assume now that (\ref{eq:22-l}) and (\ref{eq:21-l}) hold. Then clearly
(\ref{eq:19-l}) is satisfied for all $x\in\mathcal{B}$. Also, since
the following holds for $x\in\mathcal{B}^{c}$,
\begin{align*}
\mathbb{E}\left[v(X_{2})\left|X_{1}=x\right.\right] & \leq\left(1-\left(\delta-\frac{U}{v\left(x\right)}\right)\right)v(x)\\
 & \leq\left(1-\left(\delta-\delta_{1}\right)\right)v(x),
\end{align*}
it follows that (\ref{eq:20-l}) is satisfied for $x\in\mathcal{B}^{c}$ with $\delta\leftarrow\delta-\delta_{1}.$ 

To prove (\ref{eq:23-l}), write
\begin{align*}
\mathbb{E}\left[v(X_{m})\left|X_{1}=x\right.\right] & =\mathbb{E}\left[\mathbb{E}\left[v(X_{m})\left|X_{d-1},X_{1}=x\right.\right]\right]\\
 & \leq U+\left(1-\delta\right)\mathbb{E}\left[v(X_{m-1})\left|X_{1}=x\right.\right]\:{\rm by\: Markov\: property\: and}\:(\ref{eq:22-l})
\end{align*}
and hence by induction, 
\begin{align*}
\mathbb{E}\left[v(X_{m})\left|X_{1}=x\right.\right] & \leq U\sum_{i=0}^{m-1}\left(1-\delta\right)^{i}+\left(1-\delta\right)^{m}v\left(x\right)\\
 & \leq\frac{U}{\delta}+\left(1-\delta\right)^{m}v\left(x\right).
\end{align*}

\end{IEEEproof}
The next lemma states that when (\ref{eq:23}) holds, the Markov process
described in section~\ref{sub:Epoch-Based-Policy}, namely $\left\{ \left(T_{m},S_{m}\right)\right\} _{m=1}^{\infty}$
(where $T_m$ is the duration of the $m$-th epoch and $S_m$
is the system state at the end of the $m$-th epoch)
has the drift property described in Lemma \ref{lem:13}.
\begin{lem}
\label{lem:14}For the Markov process $\left\{ \left(T_{m},S_{m}\right)\right\} _{m=1}^{\infty}$
define $v\left(\left(\tau,s\right)\right)=\tau.$ If $\hat{T}\left(\boldsymbol{\lambda}\right)<1$
then there are $U>0$ and $\delta>0$ such that. 
\[
\mathbb{E}\left[v\left(\left(T_{2},S_{2}\right)\right)\left|\left(T_{1},S_{1}\right)=\left(\tau,s\right)\right.\right]\leq U+\left(1-\delta\right)v\left(\left(\tau,s\right)\right)\:{\rm for\: all\;}\left(\tau,s\right)\in\mathcal{G},
\]
and (\ref{eq:21-l}) is also satisfied.\end{lem}
\begin{IEEEproof}
Using the definition of $v,$ and the fact that given $\boldsymbol{k}_{1}$
and $S_{1}$, $T_{2}$ is independent of $T_{1},$ write, 
\begin{align}
\mathbb{E}\left[v\left(\left(T_{2},S_{2}\right)\right)\left|\left(T_{1},S_{1}\right)=\left(\tau,s\right)\right.\right] & =\mathbb{E}\left[T_{2}\left|T_{1}=\tau,\: s_{1}=s\right.\right]\nonumber \\
 & =\mathbb{E}\left[\mathbb{E}\left[T_{2}\left|T_{1}=\tau,\: s_{1}=s,\boldsymbol{k}_{1}\left(\tau\right)\right.\right]\left|\left(T_{1},S_{1}\right)=\left(\tau,s\right)\right.\right]\nonumber \\
 & =\mathbb{E}\left[\bar{T}_{s}^{\pi_{\boldsymbol{k}_{1},s}}\left(\boldsymbol{k}_{1}\left(\tau\right)\right)\right]\label{eq:25-z}
\end{align}
We have by construction of $\pi_{\epsilon},$ 
\begin{align}
\mathbb{E}\left[\bar{T}_{s}^{\pi_{\boldsymbol{k}_{1},s}}\left(\boldsymbol{k}_{1}\left(\tau\right)\right)\right] & \leq\mathbb{E}\left[\bar{T}_{s}^{*}\left(\boldsymbol{k}_{1}\left(\tau\right)\right)\right]+\epsilon\nonumber \\
 & \leq\mathbb{E}\left[\bar{T}^{*}\left(\boldsymbol{k}_{1}\left(\tau\right)\right)\right]+\epsilon.\label{eq:19-ff}
\end{align}
Since the arrival process vectors are i.i.d, it holds with probability
1, 
\[
\lim_{\tau\rightarrow\infty}\frac{\boldsymbol{k}_{1}\left(\tau\right)}{\tau}=\boldsymbol{\lambda},
\]
and,
\begin{align}
\lim_{\tau\rightarrow\infty}\frac{\bar{T}^{*}\left(\boldsymbol{k}_{1}\left(\tau\right)\right)}{\tau} & =\lim_{\tau\rightarrow\infty}\frac{\bar{T}^{*}\left(\frac{\boldsymbol{k}_{1}(\tau)}{\tau}\tau\right)}{\tau}\nonumber \\
 & =\hat{T}\left(\boldsymbol{\lambda}\right)\:{\rm by\:(\ref{eq:10-f})}\label{eq:25-1}
\end{align}
We will show at the end of the proof that the sequence $\bar{T}^{*}\left(\boldsymbol{k}_{1}\left(\tau\right)\right)/\tau,\:\tau=1,...$
is uniformly integrable, which will imply that
\begin{align*}
\limsup_{\tau\rightarrow\infty}\frac{\mathbb{E}\left[T_{2}\left|T_{1}=\tau,\: s_{n}=s\right.\right]}{\tau} & \leq\lim_{\tau\rightarrow\infty}\mathbb{E}\left[\frac{\bar{T}^{*}\left(\boldsymbol{k}_{1}\left(\tau\right)\right)}{\tau}\right]+\epsilon\quad{\rm by\;(\ref{eq:25-z}),\:(\ref{eq:19-ff})}\\
 & =\mathbb{E}\left[\lim_{\tau\rightarrow\infty}\frac{\bar{T}^{*}\left(\boldsymbol{k}_{1}\left(\tau\right)\right)}{\tau}\right]+\epsilon\:{\rm by\: uniform\: integrability}\\
 & =\hat{T}\left(\boldsymbol{\lambda}\right)+\epsilon\:{\rm by\:(\ref{eq:25-1})}.
\end{align*}
Therefore, for $\delta$ such that $0<\delta<1-\hat{T}\left(\boldsymbol{\lambda}\right)-\epsilon,$
there exists $\tau_{\delta}$ such that for all pairs $\left(\tau,s\right)$
with $\tau>\tau_{\delta}$ it holds, 
\[
\mathbb{E}\left[T_{2}\left|T_{1}=\tau,\: S_{1}=s\right.\right]\leq\left(1-\delta\right)\tau,
\]
hence, 
\[
\mathbb{E}\left[T_{2}\left|T_{1}=\tau,\: S_{1}=s\right.\right]\leq U+\left(1-\delta\right)\tau,
\]
where
\[
U=\max_{\left(\tau,s\right)\in\mathcal{G}:\tau\leq\tau_{\delta}}\mathbb{E}\left[T_{2}\left|T_{1}=\tau,\: S_{1}=s\right.\right].
\]
Also, (\ref{eq:21-l}) is satisfied since $\lim_{\tau\rightarrow\infty}v\left(\tau\right)=\tau=\infty.$

It remains to show that $\bar{T}^{*}\left(\boldsymbol{k}_{1}\left(\tau\right)\right)/\tau,\:\tau=1,2...$
is uniformly integrable. Using (\ref{eq:2-1}) we have 
\begin{equation}
0\leq\frac{\bar{T}^{*}\left(\boldsymbol{k}_{1}(\tau)\right)}{\tau}\leq C_{1}\sum_{i=1}^{n}\frac{k_{1,i}\left(\tau\right)}{\tau}+C_{0}\label{eq:25}
\end{equation}
 Now, we have with probability one, 
\[
\lim_{\tau\rightarrow\infty}\frac{k_{1,i}\left(\tau\right)}{\tau}=\lambda_{i}
\]
On the other hand, since the length of an epoch $T_{1}$ is independent
of the arrivals during this epoch, we have 
\[
\frac{\mathbb{E}\left[k_{1,i}\left(\tau\right)\right]}{\tau}=\frac{\lambda_{i}\tau}{\tau}=\lambda_{i}
\]
Since the nonnegative sequences $\frac{k_{i}\left(\tau\right)}{\tau},\: i=1,2,..,n,\:\:\tau=1,2,...$
converge both with probability one and in expectation, they are uniformly
integrable (see Theorem 16.4 in~\cite{Bi95}). Using this fact, uniform
integrability of $\bar{T}^{*}\left(\boldsymbol{k}_{1}\left(\tau\right)\right)/\tau$
follows from (\ref{eq:25}).
\end{IEEEproof}
We next present the main theorem of this section, which shows
the stability of policy $\pi_{\epsilon}.$ 
\begin{thm}
For any $\boldsymbol{\lambda}\geq\boldsymbol{0}$ such that 
\begin{equation}
\hat{T}\left(\boldsymbol{\lambda}\right)<1,\label{eq:31new}
\end{equation}
policy $\pi_{\epsilon}$ stabilizes the system. \end{thm}
\begin{IEEEproof}
The idea of the proof is the following. Assume that the system starts
at time $t=0$ in system state $s,$ with $\boldsymbol{A}(0)=\boldsymbol{k}$ packets at the inputs.
We use the queue occupancy notation
of $\boldsymbol{Q}_{s}^{\pi}(t)$, $Q_{s}^{\pi}\left(t\right)$
from Section \ref{sec:Stability---Necessity}, but we henceforth omit the indices $s$ and $\pi$ to simplify the notation.
Under $\pi_{\epsilon}$, it will be shown through Theorem \ref{thm:Meyn}
that (\ref{eq:31new}) implies that we can identify a state $\left(\tau_{a},s_{a}\right)$
to which the chain $\left(T_{m},S_{m}\right)_{m=1}^{\infty}$ returns
infinitely often. Define $m_{l},\: l=1,....,$
to be the sequence of epoch indices when the Markov chain is in
state $\left(\tau_{a},s_{a}\right)$. Then, due to the Markov property,
the process consisting of the successive intervals between the times
at which the process $\left(T_{m},S_{m}\right)_{m=1}^{\infty}$ returns
to state $\left(\tau_{a},s_{a}\right)$, i.e., 
\begin{equation}
L_{l}=\sum_{j=m_{l}+1}^{m_{l+1}}T_{j},\: l=1,2,...\label{eq:28-0}
\end{equation}
consists of i.i.d. random variables and, as will be seen,
\begin{equation}
\mathbb{E}\left[L_{l}\right]<\infty.\label{eq:finite58}
\end{equation}
Hence, the process 
\[
Z_{0}=\sum_{j=1}^{m_{1}}T_{j},\: Z_{l}=Z_{l-1}+L_{l},\: l\geq1,
\]
constitutes a (delayed) renewal process. 

Observe next that by the operation of $\pi_{\epsilon}$, $\left\{ \boldsymbol{Q}(Z_{l})\right\} _{l=0}^{\infty},$
the number of packets in system at times $Z_{l}$, is statistically
the same as the number of arrivals in a interval of length $\mbox{\ensuremath{\tau}}_{a}$.
Since packet arrivals are i.i.d and the operations of the process
during the interval $\tau_{a}$ do not depend on these arrivals, $\left\{ \boldsymbol{Q}(Z_{l})\right\} _{l=0}^{\infty},$
consists of i.i.d. random variables with $\mathbb{E}\left[\boldsymbol{Q}(Z_{l})\right]=\boldsymbol{\lambda}\tau_{a}<\infty$.
This, and the operation of $\pi_{\epsilon}$ imply that the process
$\left\{ \boldsymbol{Q}(t)\right\} _{t=0}^{\infty}$, is regenerative
with respect to $\left\{ Z_{l}\right\} _{l=0}^{\infty}.$ Let $g$
be the period of the distribution of the cycle length $L_{l}$. It
then follows from Corollary 1.5 p.128 in \cite{asmussen}
and (\ref{eq:finite58}) that
\begin{align}
\lim_{\alpha\rightarrow\infty}\frac{1}{g}\sum_{\beta=0}^{g-1}\Pr\left(Q\left(\alpha{g}+\beta\right)>q\right) & =\lim_{\alpha\rightarrow\infty}\frac{1}{g}\sum_{\beta=0}^{g-1}\mathbb{E}\left[1_{\left\{ Q\left(\alpha{g}+\beta\right)>q\right\} }\right]\nonumber \\
 & =\frac{\mathbb{E}\left[\sum_{j=0}^{L_{1}-1}1_{\left\{ Q\left(Z_{0}+j\right)>q\right\} }\right]}{\mathbb{E}\left[L_{1}\right]}\label{eq:33new}
\end{align}
Observe next that the random variables $Y_{j}(q)=1_{\left\{ Q\left(Z_{0}+j\right)>q\right\} }$
are decreasing in $q,$ and since $Q\left(Z_{0}+j\right)$ are finite,
$\lim_{q\rightarrow\infty}1_{\left\{ Q\left(Z_{0}+j\right)>q\right\} }=0.$
Using the monotone convergence theorem we then have, 
\begin{align}
\lim_{q\rightarrow\infty}\mathbb{E}\left[\sum_{j=0}^{L_{1}-1}1_{\left\{ Q\left(Z_{0}+j\right)>q\right\} }\right] & =\mathbb{E}\left[\lim_{q\rightarrow\infty}\sum_{j=0}^{L_{1}-1}1_{\left\{ Q\left(Z_{0}+j\right)>q\right\} }\right]\nonumber \\
 & =0.\label{eq:34new}
\end{align}
From (\ref{eq:33new}), (\ref{eq:34new}) we conclude that 
\begin{equation}
\lim_{q\rightarrow\infty}\lim_{alpha\rightarrow\infty}\sum_{\beta=0}^{g-1}\Pr\left(Q\left(\alpha{g}+\beta\right)>q\right)=0\label{eq:35new}
\end{equation}
 Since for $t=\alpha_{t}g+\beta_{t}$ it holds 
\[
\Pr\left(Q\left(t\right)>q\right)\leq\sum_{\beta=0}^{g-1}\Pr\left(Q\left(\alpha_{t}g+\beta\right)>q\right)
\]
 we conclude from (\ref{eq:35new}) that 
\[
\lim_{q\rightarrow\infty}\limsup_{t\rightarrow\infty}\Pr\left(Q\left(t\right)>q\right)=0
\]
i.e., policy $\pi_{\epsilon}$ is stable. 

To implement the plan outlined above we must show the existence of
a state to which the Markov chain returns infinitely often, as well
as (\ref{eq:35new}). For this we will use Theorem \ref{thm:Meyn}
but because of the generality of the model under consideration, we
cannot apriori claim irreducibility and aperiodicity in order to apply
it directly. Instead, we rely first on Lemma \ref{lem:13} using the
result of Lemma \ref{lem:14}. 

Let ${C}\left(\left(\tau,s\right)\right)$ be the communicating
class to which a state $\left(\tau,s\right)$ belongs. We consider
two cases as follows. 

a) If ${C}\left(\left(\tau,s\right)\right)$ is essential,
and $\left(T_{1},\: S_{1}\right)=\left(\tau,s\right),$ we have $\left(T_{m},\: S_{m}\right)\in{C}\left(\left(\tau,s\right)\right)\: m=1,2,...$
and the evolution of the process with initial condition $\left(\tau,s\right)$
constitutes an irreducible Markov Chain. If this chain is periodic
with period $d$, then the process $\left\{ \left(T_{dk+1},\: S_{dk+1}\right)\right\} _{k=0}^{\infty}$
is an aperiodic Markov Chain, \cite{chung} page 14. For
this chain, we can apply Theorem \ref{thm:Meyn} to show positive
recurrence, as follows. Since by Lemma \ref{lem:14} the process $\left\{ \left(T_{m},S_{m}\right)\right\} _{m=1}^{\infty}$
satisfies (\ref{eq:22-l}), it also satisfies (\ref{eq:23-l}). Hence
the process $\left\{ \left(T_{dk+1},\: S_{dk+1}\right)\right\} _{k=0}^{\infty}$
satisfies (\ref{eq:22-l}) and since $\lim_{\tau\rightarrow\infty}v\left(\left(\tau,s\right)\right)=\infty$,
it also satisfies (\ref{eq:21-l}). Therefore, by Lemma \ref{lem:13}
we can apply Theorem \ref{thm:Meyn} to $\left\{ \left(T_{dk+1},\: S_{dk+1}\right)\right\} _{k=0}^{\infty}$
to deduce that it is geometrically ergodic with 
\begin{equation}
\mathbb{E}\left[v\left(\overset{\smallfrown}{X}\right)\right]=\mathbb{E}\left[\overset{\smallfrown}{T}\right]<\infty.\label{eq:28i}
\end{equation}
From the above discussion we conclude that the initial state $\left(\tau,s\right)$
is visited infinitely often, and the successive visit indices are
of the form $m_{l}=dV_{l}+1,\: l=0,1,2...,\: V_{1}=0,$ where $V_{l},\: l=1,...$
are integer valued i.i.d. random variables with 
\begin{equation}
\mathbb{E}\left[V_{l}\right]<\infty\label{eq:30-w}
\end{equation}
Let now,
\begin{equation}
\tilde{L}_{k}=\sum_{j=1}^{d}T_{dk+j}\: k=0,1,...\label{eq:38new}
\end{equation}
The nonnegative process $\left\{ \tilde{L}_{k}\right\} _{k=0}^{\infty}$
is regenerative with respect to $\left\{ V_{l}\right\} _{l=1}^{\infty}$
and by the regenerative theorem and (\ref{eq:30-w}) it holds, 
\begin{equation}
\lim_{k\rightarrow\infty}\frac{\mathbb{E}\left[\sum_{m=0}^{k}\tilde{L}_{m}\right]}{k}=\frac{\mathbb{E}\left[\sum_{m=0}^{V_{1}-1}\tilde{L}_{m}\right]}{\mathbb{E}\left[V_{1}\right]}.\label{eq:lim65}
\end{equation}
Observe next that by (\ref{eq:28-0}) and (\ref{eq:38new}), 
\begin{equation}
\mathbb{E}\left[\sum_{m=0}^{V_{1}-1}\tilde{L}_{m}\right]=\mathbb{E}\left[L_{1}\right]\label{eq:equiv65}
\end{equation}
 Hence in order to show (\ref{eq:finite58}) it suffices to show 
\begin{equation}
\mathbb{E}\left[\sum_{m=0}^{V_{1}-1}\tilde{L}_{m}\right]<\infty,\label{eq:equiv66}
\end{equation}
or, by (\ref{eq:lim65}),
\begin{equation}
\lim_{k\rightarrow\infty}\frac{\mathbb{E}\left[\sum_{m=0}^{k}\tilde{L}_{m}\right]}{k}<\infty.\label{eq:finally}
\end{equation}
Notice that by (\ref{eq:38new}) we have, 
\begin{align*}
\mathbb{E}\left[\sum_{m=0}^{k}\tilde{L}_{m}\right] & =\mathbb{E}\left[\sum_{m=0}^{d(k+1)}T_{m}\right]\\
 & =\sum_{m=0}^{d(k+1)}\mathbb{E}\left[T_{m}\right]\\
 & \leq\sum_{m=0}^{d(k+1)}\left(\frac{U}{\delta}+\left(1-\delta\right)^{m}\tau\right)\mbox{ by (}\ref{eq:23-l})\\
 & \leq\frac{U}{\delta}\left(dk+d+1)\right)+\frac{\tau}{\delta}
\end{align*}
from which (\ref{eq:finally}) follows.

b) Consider next the case where ${C}\left(\left(\tau,s\right)\right)$
is inessential, i.e., there is at least one state $y\in\mathcal{G}-{C}\left(\left(\tau,s\right)\right)$
such that for $x\in{C}\left(\left(\tau,s\right)\right),$
$x\rightsquigarrow y$ but $y\not\rightsquigarrow x$; here, with
$x,$ $y$ we denote pairs of the form $(\tau,s).$ Hence there is
at least one other communicating class reachable from ${C}\left(\left(\tau,s\right)\right).$
The communicating classes reachable from ${C}\left(\left(\tau,s\right)\right)$
will be either essential or inessential. We argue that the process
$\left\{ \left(T_{m},\: S_{m}\right)\right\} _{m=1}^{\infty}$ will
enter an essential class in finite time. Assume the contrary, that
is, there is a set of sample paths $\Omega_{I}$ with $\Pr\left\{ \Omega_{I}\right\} >0$,
for which the process remains always in some inessential class. Since
inessential states are nonrecurrent (see~\cite{chung}, Theorem 4, p. 18)
the process visits each inessential state only a finite number of
times. This implies that on $\Omega_{I},$ $\lim_{m\rightarrow\infty}T_{m}=\infty,$
and since $\Pr\left\{ \Omega_{l}\right\} >0,$ we conclude that 
\[
\mathbb{E}\left[\lim_{m\rightarrow\infty}T_{m}\left|(T_{1},S_{1})=\left(\tau,s\right)\right.\right]=\infty.
\]
Applying next Fatou's Lemma we have 
\begin{align*}
\liminf_{m\rightarrow\infty}\mathbb{E}\left[T_{m}\left|(T_{1},S_{1})=\left(\tau,s\right)\right.\right] & \geq\mathbb{E}\left[\lim_{m\rightarrow\infty}T_{m}\left|(T_{1},S_{1})=\left(\tau,s\right)\right.\right]\\
 & =\infty,
\end{align*}
which contradicts (\ref{eq:23-l}). 

Since the process enters again an essential class in finite time,
we can apply the arguments of case a) to complete the proof. 
\end{IEEEproof}

\end{document}